\documentclass[final,onecolumn] {elsarticle}
\usepackage{amsmath,latexsym,amssymb,setspace,stackrel,tikz,graphicx,caption,color,epstopdf,subfig,epstopdf,pgfplots,url}
\usepackage{bbm,bm}
\long\def\comment#1{}

\newfont{\bbb}{msbm10 scaled 700}

\newfont{\bb}{msbm10 scaled 1100}

\newcommand{\zerov}{{\bf 0}}

\newcommand{\Fm}{{\bf F}}

\newcommand{\Hm}{{\bf H}}
\newcommand{\Id}{{\bf I}}

\newcommand{\Sm}{{\bf S}}

\newcommand{\Ac}{{\cal A}}

\newcommand{\Cc}{{\cal C}}
\newcommand{\Dc}{{\cal D}}
\newcommand{\Ec}{{\cal E}}

\newcommand{\Ic}{{\cal I}}

\newcommand{\Nc}{{\cal N}}

\newcommand{\Uc}{{\cal U}}
\newcommand{\Wc}{{\cal W}}

\newcommand{\Xc}{{\cal X}}
\newcommand{\Yc}{{\cal Y}}
\newcommand{\Zc}{{\cal Z}}

\newcommand{\Lambdam}{\hbox{\boldmath$\Lambda$}}
\newcommand{\Deltam}{\hbox{\boldmath$\Delta$}}

\newcommand{\Psim}{\hbox{\boldmath$\Psi$}}

\newcommand{\Thetam}{\hbox{\boldmath$\Theta$}}
\newcommand{\thetam}{\hbox{\boldmath$\theta$}}

\renewcommand{\arg}{{\hbox{arg}}}

\newcommand{\herm}{{\sf H}}

\newcommand{\transp}{{\sf T}}

\newcommand{\Afd}{\mbox{$\boldsymbol{\mathcal{A}}$}}

\newcommand{\Cfd}{\mbox{$\boldsymbol{\mathcal{C}}$}}
\newcommand{\Dfd}{\mbox{$\boldsymbol{\mathcal{D}}$}}
\newcommand{\Efd}{\mbox{$\boldsymbol{\mathcal{E}}$}}

\newcommand{\Xfd}{\mbox{$\boldsymbol{\mathcal{X}}$}}
\newcommand{\Yfd}{\mbox{$\boldsymbol{\mathcal{Y}}$}}
\newcommand{\Zfd}{\mbox{$\boldsymbol{\mathcal{Z}}$}}

\newcommand{\cb}{{\bm{c}}}

\newcommand{\hb}{{\bm{h}}}

\newcommand{\wb}{{\bm{w}}}
\newcommand{\xb}{{\bm{x}}}
\newcommand{\yb}{{\bm{y}}}
\newcommand{\zb}{{\bm{z}}}

\biboptions{sort&compress}
\usetikzlibrary{spy}
\begin{document}
\title{\huge Nonlinear Distortion Reduction in OFDM from Reliable Perturbations in Data Carriers}
\author[USC]{Ebrahim~B.~Al-Safadi}
\ead{alsafadi@usc.edu}
\author[KAUST,KFUPM]{Tareq~Y.~Al-Naffouri\corref{cor1}}
\ead{tareq.alnaffouri@kaust.edu.sa}
\author[KAUST]{Mudassir~Masood}
\ead{mudassir.masood@kaust.edu.sa}
\author[KAUST]{Anum~Ali}
\ead{anum.ali@kaust.edu.sa}
\address[USC]{University of Southern California, Los Angeles, CA, United States.}
\address[KAUST]{King Abdullah University of Science \& Technology, Thuwal, Saudi Arabia.}
\address[KFUPM]{King Fahd University of Petroleum \& Minerals, Dhahran, Saudi Arabia.}
\cortext[cor1]{Corresponding author. Tel./fax: +966-54-470-0795}
\tnotetext[t1]{This work was supported by the Deanship of Scientific Research at King Fahd University of Petroleum \& Minerals (KFUPM) under Research Grant FT100030 and by the Viterbi fellowship from the Graduate School at the University of Southern California.}
\tnotetext[t2]{Part of this work has appeared in the $13^{th}$ IEEE International Workshop on Signal Processing Advances in Wireless Communications (SPAWC 2012) \cite{Safadi_SPAWC}.}
%
%\markboth{Signal Processing. SIGPRO-D-**-*****}{}

\begin{abstract}
A novel method for correcting the effect of nonlinear distortion in orthogonal frequency division multiplexing signals is proposed. The method depends on adaptively selecting the distortion over a \emph{subset} of the data carriers, and then using tools from compressed sensing and sparse Bayesian recovery to estimate the distortion over the other carriers. Central to this method is the fact that carriers (or tones) are decoded with different levels of confidence, depending on a coupled function of the magnitude and phase of the distortion over each carrier, in addition to the respective channel strength. Moreover, as no pilots are required by this method, a significant improvement in terms of achievable rate can be achieved relative to previous work.
\end{abstract}
\maketitle
\section{Introduction}
Multicarrier signaling schemes such as orthogonal frequency division multiplexing (OFDM) are highly susceptible to nonlinear distortion at all stages of the transmission process. This is partly due to the impulsive nature of these signals in the time domain, where the superposition of modulated waveforms takes place. When nonlinear distortion is confined to the transmitter, many proposed methods in the literature use reserved carriers (or tones) to carry information about this distortion to the receiver, at the obvious cost of reducing data-rate \cite{Tellado_Book,Chen_tone,Tone_reservation_new,active_set}. The central idea in these techniques is to construct clipping (or peak-reducing) signals by performing a constrained search at the transmitter, one which confines the frequency support of the clipping signal to the reserved carriers, while reducing the peaks of the data signal in the time domain. These approaches are not generally robust, as they demand that the frequency support of the data and clipping signals remain strictly disjoint throughout the transmission process, and add significant complexity at the transmitter.

To combat this, techniques based on compressive sensing (CS) that were tuned to clipped OFDM models were proposed in \cite{Safadi2}. These techniques removed the need for any constrained search at the transmitter, since the receiver could detect the entire clipping signal by observing a subset of its frequency components available on the reserved tones, provided that the signal is sparse in time. Consequently, the need to maintain orthogonality in frequency was completely relaxed, but the need for a significant amount of reserved carriers persisted. To avoid this loss in data-rate, the authors in \cite{Mohammadnia} proposed using the channel estimation pilots for this purpose. Nonetheless, this approach severely limits the number of measurements available to the CS algorithm and hence its ability to deal with severe clipping scenarios. It also does not make use of available information such as clipping likelihood and phase resemblance in the time domain (i.e., phase resemblance between the clipped and clipping signal) as done in \cite{Safadi2}.

In this paper, a fundamentally different approach to these methods is pursued. Specifically, in contrast to the authors' previous work on the topic of using CS concepts in OFDM \cite{Safadi2}, the technique presented in this paper does not require any orthogonality between the frequency support of the data and the distortion, and no tones (null, edge, channel pilots, or otherwise) are needed either. The receiver is free to select which and how many data tones it will use to read off differential observations, and will use them to estimate and cancel the entire distortion over the tones. In addition, no data-rate is lost by employing the proposed strategy. Furthermore, a significant tradeoff also exists in regard to complexity, distortion tolerance, and robustness to channel estimation errors, so that the user has many algorithms to use within the proposed framework. Similarly, in contrast to another recent work \cite{Safadi_SPAWC}, the current paper introduces an entirely new and rigorous way of analyzing the reliability of tones. In addition, it also introduces a method to finetune the performance of CS by minimizing the probability of incorrect measurements and maximizing the clipping-to-noise ratio (CNR).

This framework is made possible by \emph{jointly} taking three phenomena into account. The first is that not all tones carry correct decoding information to the receiver. The second is that the receiver can probabilistically assign levels of confidence to each tone, and the third is that the distortion is sparse in the time domain. These phenomena motivate us to employ CS and sparse recovery techniques much more effectively compared to previous techniques, as we can quantify the perturbations on data tones, select the most reliable ones, and then most-importantly, use the power of CS techniques to recover the significant time domain distortions. Our major contributions include:
\begin{enumerate}
  \item Formulating CS models within a pilotless transmission framework.
  \item Proposing systematic methods to adaptively select the tone subset to sense over, among the combinatorially large possibilities of pilotless CS models.
  \item Developing a novel method for accurately assessing the reliability of estimated coefficients based on their symbol-wise magnitudes, phases, and relative locations to other constellation points, as well as channel strengths. %This reliability computation is in the form of a \emph{vector}-wise likelihood ratio, unlike the bit-wise likelihood ratio used in the literature which cannot be easily approximated by geometric profiles.
  \item Deriving a closed-form expression that characterizes the modes of behavior of the reliability function, and devising geometrically-inspired approximations based on this expression for quick and efficient selection of the tone subset over each OFDM block.
  \item Proposing dual-stage construction of the tone subset, where the first stage minimizes the probability of incorrect measurements, while the second maximizes an CNR metric to optimize CS performance.
  \item Providing probabilistic upper bounds for choosing the \emph{number} of tones in the CS model without risking incorrect measurements.
\end{enumerate}
The remainder of the paper is organized as follows. Section \ref{trans_model} briefly describes the transmission and distortion models. Section \ref{cs_model_sec} demonstrates how a pilotless CS model can be derived within the previous transmission model. Subsequently, Section \ref{tone_selection_criteria}, the heart of the paper, focuses on selecting the subset of tones used for CS. This includes developing reliability assessment criteria, deriving analytical approximations for quick and efficient assessment, selecting the number of tones, establishing dual-stage subset selection to maximize CS performance, and, finally, condensing the major results into an algorithm. Section \ref{sim_results} presents our simulations and Section \ref{conclusion} concludes the paper.
\subsection{Notation}
We use \textit{regular font} for scalars and \textit{boldface} letters for matrices and vectors. To distinguish between vectors in the time and frequency domains, we use \textit{boldface calligraphic} notation for vectors in the frequency domain (e.g. $\Xfd, \underline{\Xfd},\Yfd$) and \textit{boldface lowercase} letters for their corresponding time domain representations (e.g. $\xb, \underline{\xb},\yb$). We use $\Xc(k)$ to denote the $k$th coefficient of $\Xfd$, or more simply $\Xc$, when it is clear from the context. Moreover, we use $\Xfd_\Omega$ to represent a vector formed by selecting the coefficients of $\Xfd$ indexed by set $\Omega$. Similarly, $\Yfd_\Omega$ is the vector formed by indexing the corresponding elements of vector $\Yfd_\Omega$ according to the index set $\Omega$. We further define $\Sm_\Omega$ to be a diagonal binary selection matrix, with $|\Omega|$ number of $1$'s at locations along its diagonal specified by the tone set $\Omega$.
\section{Transmission and Clipping Model}\label{trans_model}
In an OFDM system, serially incoming bits are mapped into an
$M$-ary QAM alphabet
$\Afd\!=\!\{\Ac_{0},\Ac_{1},\ldots,\Ac_{M-1}\}$ and
concatenated to form an $N$-dimensional data vector, $\Xfd=[\Xc(0) \Xc(1),\cdots,\Xc(N-1)]^\transp\in\Afd^{N}$. The time domain signal $\xb$ is obtained by an IFFT operation such that $\xb=\Fm^\herm\Xfd$, where
\begin{equation*}
F_{k}(l)=N^{-1/2}\, e^{-\jmath 2\pi kl/N},\quad k,l\in{0,1,\ldots,N-1}.
\end{equation*}
Since $\xb$ has a high PAPR, the digital samples are subject to a magnitude limiter that saturates its operands to a value of $\gamma$. Hence, instead of $\xb$, we feed $\underline{\xb}$ to the power amplifier, where
\begin{align}
\underline{x}(i) =
\begin{cases}
\gamma e^{\jmath \theta_{x(i)}}&\text{if}~|x(i)|>\gamma, \\
x(i)& \text{otherwise},
\end{cases}
\end{align}
and where $\theta_{x(i)}$ is the phase of $x(i)$. This hard-limiting operation can be conveniently thought of as adding a peak-reducing signal $\cb$ to $\xb$ so that its low-PAPR counterpart $\underline{\xb}=\xb+\cb$ is
transmitted instead. Furthermore, by setting a typical
clipping threshold, $\gamma$, on $\xb$, $\cb$ is controllably sparse in
time by the impulsive nature of $\xb$, and dense in frequency by the
uncertainty principle. We denote the temporal support of $\cb$ by
$\Ic_{\cb}=\{i:c(i)\neq0\}$ and always maintain the practical
assumption that $|\Ic_{\cb}|\ll N$.

Subsequently, $\underline{\xb}$ is convolved with a channel of impulse response $\hb\sim\Cc\Nc(\zerov,\sigma_{\hb}^{2}\Id_{L_{\hb}})$, and subjected to additive white Gaussian noise (AWGN) $\zb\sim\Cc\Nc(\zerov,\sigma_{\zb}^{2}\Id_{N})$, where $L_h$ is the length of channel impulse response. Equivalently, in the frequency domain, this translates to transmitting
\begin{align}\label{eq:freqsigmodel}
\underline{\Xfd}\!=\!\Xfd+\Cfd,
\end{align}
with complex coefficients that are now randomly pre-perturbed from the lattice $\Afd^{N}$, followed by additional multiplicative perturbations by the channel $\Hm$ and additive perturbations by the noise  $\Zfd\sim\Cc\Nc(\zerov,\sigma_{\bm{z}}^2\Id_{N})$ at the receiver. By virtue of the added cyclic prefix (of length $>L_{\hb}$), the circulant channel matrix $\Hm$ can be decomposed and expressed as $\Hm\!=\!\Fm^\herm\Lambdam\Fm$ where $\Lambdam$ is an $N\times N$ diagonal matrix composed of the frequency-domain channel gains, $\{\lambda(k)\}_{k=1}^{N}$. As a result the frequency domain received signal reads $\Yfd=\Lambdam\underline{\Xfd}+\Zfd$, where, for the moment, we make the practical assumption that the channel
coefficients are known at the receiver. Consequently, $\underline{\Xfd}$ can be directly recovered scalar-wise from $\bm{\Yfd}$, i.e.,
\begin{eqnarray}\label{X_bar_estimate}
\hat{\underline{\Xc}}(k)=\lambda^{-1}(k)\Yc(k)=\Xc(k)+\Cc(k)+\lambda^{-1}(k)\Zc(k),
\end{eqnarray}
where we use the notation $\hat{\Xc}$ to represent the equalized estimate of $\Xc$ at the receiver. Writing (\ref{X_bar_estimate}) in vector notation yields
\begin{align}\label{eq:vecXhat}
\hat{\underline{\Xfd}} = \Xfd + \Cfd + \Lambdam^{-1}\Zfd.
\end{align}
Let $\Dc(k)\triangleq \Cc(k)+\lambda^{-1}(k)\Zc(k)$ denote the general distortion on $\Xc(k)$, We let $f_\Dc$ to be the pdf of the general distortion $\Dc$ which we assume to be zero mean circularly symmetric Gaussian with variance $\sigma_\Dc^2$. (Please refer to Appendix A for details regarding the derivation of $\sigma_\Dc^2$.)  Equation (\ref{eq:vecXhat}) could now be written as
\begin{align}\label{eq:xhateqxplusd}
\hat{\underline{\Xfd}} = \Xfd + \Dfd.
\end{align}
Treating the clipping distortion as additive noise, an maximum likelihood (ML) decoder will recover $\Xc(k)$ by simply mapping $\hat{\underline{\Xc}}(k)$ to the nearest
constellation point\footnote{While $\Ac_m$ refers to the $m$th constellation point $(1\!\le\!m\!\le\!M)$, we reserve $\langle \hat{\Xc}(k) \rangle$ to denote the nearest constellation point corresponding to the $k$th received data sample $\hat{\Xc}(k)$. Furthermore, note that the \emph{true} constellation point corresponding to $\hat{\Xc}(k)$ is $\Xc(k)$.} $\langle \hat{\underline{\Xc}}(k) \rangle$, where $\langle\hat{\underline{\Xc}}(k)\rangle\triangleq
\arg\min_{\Ac_{m}\in\bm{\Ac}}|\hat{\underline{\Xc}}(k)-\Ac_{m}|$.  In other words, the operation $\langle \hat{\Xc}(k) \rangle$ corresponds to rounding $\hat{\Xc}(k)$ to the nearest constellation point. Note that $\langle \hat{\Xc}(k) \rangle$ does not need to be the true constellation point, i.e., $\langle \hat{\Xc}(k) \rangle$ might be different from $\Xc(k)$. Such a hard-decoding scheme is very efficient in the classical AWGN scenario for high signal-to-noise ratio (SNR). However, in our case, in addition to the additive noise $\lambda^{-1}(k)\Zc(k)$, we have a $\gamma$-dependent source of perturbation $\Cc(k)$ which is independent of the SNR. CS and similar sparse recovery algorithms seem to be a very sensible solution towards recovery of $\Cc(k)$. Since $\cb$ is sparse in the time domain, a \textit{partial} observation of $\cb$ in frequency domain is sufficient to estimate $\cb$ and hence $\Cfd$ in one shot. This would certainly get around the problem of unreliable perturbations as CS algorithms, for instance, can be totally blind to them and still offer near optimal signal reconstruction under mild conditions \cite{Candes1}. The main issue is to decide which partial observation to use. This will be the topic of the following section.
\section{Development of Compressive Sensing Models with No Tone Reservation}\label{cs_model_sec}
With the addition of the general distortion vector $\Dfd$ to the data vector $\Xfd$, we expect that part of the data samples will be severely perturbed such that they fall out of their true decision regions. Let $\langle\Xc(k)\rangle$ denote the decoded data sample corresponding to $\Xc(k)$, then the true decision region for $\Xc(k)$ is defined as $\mathbb{Q}(k)\!\triangleq\!\{\,\Xc(k)+\Uc\!\in\!\mathbb{C}\!:\!\langle\Xc(k)+\Uc\rangle\!=\!\Xc(k)\}$ where  $\Uc$ is a factor which when added to $\Xc(k)$ keeps it in its true decision region. Moreover, denote by $\Omega_{T}\!=\!\{k\in\Omega\!:\!\langle\Xc(k)+\Dc(k)\rangle\!=\!\Xc(k)\}$ the subset of data tones in $\Omega=\{1,2,\ldots,N\}$ in which the perturbations do not cause data samples to cross their true decision regions. Let  $\bar{\Omega}_T=\Omega\backslash\Omega_{T}$ be its complement. Over the data tones of $\Omega_{T}$, the equality in $\langle \underline{\Xc}(k)\rangle=\Xc(k)$ is true and hence from (\ref{eq:freqsigmodel}) and (\ref{eq:xhateqxplusd}), $\Dfd_{\Omega_{T}}=\underline{\hat{\Xfd}}_{\Omega_{T}}-\Xfd_{\Omega_T}=\underline{\hat{\Xfd}}_{\Omega_{T}}-\langle\underline{\hat{\Xfd}}_{\Omega_{T}}\rangle$. This is not true at the complement set $\bar{\Omega}_T$ at which $\Dfd_{\bar{\Omega}_{T}} = \underline{\hat{\Xfd}}_{\bar{\Omega}_{T}}-\Xfd_{\bar{\Omega}_{T}} \ne \underline{\hat{\Xfd}}_{\bar{\Omega}_{T}}-\langle \hat{\Xfd}_{\bar{\Omega}_{T}} \rangle$. More generally, we can write
\begin{eqnarray}
\Dfd=\Sm_{\Omega_{T}}\left(\underline{\hat{\Xfd}}-\langle
\underline{\hat{\Xfd}}\rangle\right)+\Sm_{\bar{\Omega}_{T}}\left(\underline{\hat{\Xfd}}-\Xfd\right),
\end{eqnarray}
where $\Sm_{\Omega_{T}}$ is an $N\!\times \!N$ diagonal binary selection matrix, with $|\Omega_{T}|$ number of $1$'s at locations along its diagonal specified by the tone set $\Omega_T$. It extracts the elements of the vector $\underline{\hat{\Xfd}}-\langle
\underline{\hat{\Xfd}}\rangle$ according to the tone set $\Omega_{T}$ while nulling the others. The matrix $\Sm_{\bar{\Omega}_{T}}$ is similarly defined with $1$'s along the diagonal specified by the set $\bar{\Omega}_T$. It is easy to see that $\Sm_{\Omega_{T}}\Sm_{\bar{\Omega}_{T}}=\zerov$. Practically speaking, $\Omega_{T}$ constitutes the larger part of the tone set $\Omega$. An essential part of OFDM signal recovery obviously constitutes finding $\Omega_T$ and correcting the distortions over $\bar{\Omega}_{T}$ to finally reach the state $\Omega_{T}=\Omega$.

From (\ref{eq:xhateqxplusd}) we have at the receiver,
\begin{align}\label{eq:analog_estimate}
\underline{\hat{\Xfd}} = \Xfd + \Dfd,
\end{align}
which is the \emph{analog} estimate $\underline{\hat{\Xfd}}$ of the data vector $\Xfd$ affected by the distortion $\Dfd$. We define $\Efd\triangleq \Xfd-\langle\underline{\hat{\Xfd}}\rangle$ to be a vector that is nonzero at locations where the decoded estimate at the receiver $\langle\underline{\hat{\Xfd}}\rangle$ differs from the data vector $\Xfd$. From the discussion above, we see that
\begin{align}
\underline{\hat{\Xc}}(k) - \langle \underline{\hat{\Xc}}(k) \rangle&=
    \begin{cases}
      \Dc(k), & \text{if}~\langle\underline{\hat{\Xc}}(k)\rangle=\Xc(k) \\
      \Dc(k)+\Ec(k),&\text{if}~\langle\underline{\hat{\Xc}}(k)\rangle\ne\Xc(k)
    \end{cases}
\end{align}
which allows us to write
\begin{eqnarray}
\Sm_{\Omega_{T}}(\underline{\hat{\Xfd}}-\langle\underline{\hat{\Xfd}}\rangle)=\Sm_{\Omega_{T}}\Dfd \label{difference_equivalence}
\end{eqnarray}

Note that we do not require all of $\Omega_{T}$ to recover $\cb$. Rather, we only require an arbitrary \emph{subset} $\Omega_{m}\subseteq\Omega_{T}\subseteq\Omega$ of
cardinality $m\triangleq|\Omega_{m}|\le|\Omega_{T}|$ to correctly recover $\cb$ by CS ($m$ and $|\Omega_{m}|$ will be used interchangeably as appropriate to denote the number of measurements). As a result, we can replace the equation above with $\Sm_{\Omega_{m}}(\underline{\hat{\Xfd}}-\langle\underline{\hat{\Xfd}}\rangle)=\Sm_{\Omega_{m}}\Dfd=\Sm_{\Omega_{m}}\Fm\cb+\Sm_{\Omega_{m}}\Lambdam^{-1}\Zfd$, where $\Sm_{\Omega_m}$ is also a $N\!\times\! N$ diagonal binary selection matrix defined in a similar way as $\Sm_{\Omega_T}$ in the above. We write the above equation simply as $\Yfd^\prime\triangleq\Psim\cb+\Zfd^\prime$, where $\Psim\triangleq\Sm_{\Omega_{m}}\Fm$,
$\Zfd^\prime\triangleq\Sm_{\Omega_{m}}\Lambdam^{-1}\Zfd$, and $\Yfd^\prime\triangleq\Sm_{\Omega_{m}}(\underline{\hat{\Xfd}}-\langle\underline{\hat{\Xfd}}\rangle)$ which denotes the observation vector of the differences over the tones in $\Omega_{m}$, nulled at the discarded measurements. This leads us to a pilotless CS model\footnote{The reason we stress that the CS model does not reduce transmission rate is that there have been previous alternate attempts by the authors \cite{Safadi2} and others \cite{Mohammadnia} to use compressive sensing in a tone-reservation setting which required significant reduction in data-rate.}
\begin{eqnarray}\label{lossless_CS_model}
\Yfd^\prime_{\Omega_{m}}=\Psim_{\Omega_{m}} \cb+\Zfd^\prime_{\Omega_{m}}.
\end{eqnarray}
where $\Yfd'_{\Omega_{m}}$ is the $|\Omega_{m}|$-dimensional vector
composed of the nonzero coefficients in $\Yfd^\prime$.

By inspecting (\ref{lossless_CS_model}), we notice that $\cb$ is an $N$-dimensional sparse vector in the time domain, corresponding
to the difference between the time representations of the OFDM signal $\xb$ and its clipped counterpart $\underline{\xb}$. The matrix $\Psim_{\Omega_{m}}\!\!\in\!\mathbb{C}^{m\times N}$ is obtained by $m$ random row extractions from the $N\!\times \!N$ Fourier matrix according to $\Omega_{m}$ (the cause of randomness is discussed later). The $m$-dimensional vector $\Yfd^\prime_{\Omega_{m}}$ is the corresponding partial frequency-domain observation that we use to estimate $\cb$, contaminated by the Gaussian noise vector, $\Zfd^\prime_{\Omega_{m}}$.

This is a standard model in CS \cite{Candes1,Candes2}. Note however that the parametrization by $\Omega_{m}$ actually defines a huge set of $2^{N}$ possible models\footnote{The reason is that we do not know $\Omega_m$ or even $|\Omega_m|$ and so we might have to search over all the subsets of $N$ tones giving us a total of $2^N$ models like (\ref{lossless_CS_model}) to choose from.}. In the forthcoming sections we will discuss in detail how to determine a proper model from all these possibilities. For the time being, we assume that an appropriate $\Omega_{m}$ is chosen,
and $\cb$ could therefore be recovered using any CS technique, be it convex programming, greedy pursuit, or iterative thresholding, and a very flexible region for tradeoff exists in regard to the performance and complexity of these techniques.

In this paper, we use two different schemes of CS to recover $\cb$ from
the developed CS model in (\ref{lossless_CS_model}), one from the convex relaxation group and the other from greedy pursuit methods. More specifically, the first is an adaptation of the least absolute shrinkage and selection operator (LASSO) \cite{Tibshirani} to this problem called the weighted and phase-augmented LASSO (WPAL) \cite{Safadi2}. It incorporates phase information and clipping likelihood available from the data in the time-domain to improve distortion recovery performed in the frequency-domain. Specifically, we know that $\cb$ is composed of just the clipped portions of the transmitted signal $\underline{\xb}$, so at clipping locations $\angle{c(k)} = -\angle{\underline{x}(k)}$. We also know that the closer $|\underline{x}(k)|$ to the value of the clipping threshold $\gamma$, the higher the likelihood that $\cb$ had an active coefficient at $k$. This additional information is incorporated in the CS algorithm in the form of weighting to improve its performance. Therefore, we define $\wb\triangleq||\underline{\hat{\xb}}|-\gamma|^\transp$ to be such a weighting vector to the $\ell_{1}$-norm of $\cb$ in the LASSO where $\underline{\hat{\xb}}$ refers to the estimated received clipped signal. We further define the diagonal phase matrix $\hat{\Theta}_{c}\!=\!-\exp\left(\textmd{diag}(\jmath\thetam_{\underline{\hat{\xb}}})\right)$ such that $\cb^{\textmd{WPAL}}=\hat{\Theta}_{c}|\cb^{\textmd{WPAL}}|$. With these two variables defined, the optimization problem we solve becomes
\begin{eqnarray}
\nonumber&&\!\!\!\!\!\!\!\!\!\!\!\!|\,\cb^{\textmd{WPAL}}\,|=\underset{|\cb|\in \mathbb{R}^{N}}{\arg\min}\:\:\wb^\transp|\cb|\\
\textmd{s.t.}&&\!\!\!\!\!\!\!\!\|\, \Yfd^\prime_{\Omega_{m}}-\Psim_{\Omega_{m}}\hat{\Thetam}_{\cb}
|\cb|\,\|_{2}^{2}<\epsilon \label{LASSO}
\end{eqnarray}
for some noise-dependent parameter $\epsilon$. The other technique is the fast Bayesian matching pursuit (FBMP) by Schniter \textit{et al.}
\cite{Schniter2} chosen for its superior performance and efficiency
when a relatively large number of measurements is available, which is indeed the case compared to tone reservation cases proposed in \cite{Safadi2} and \cite{Mohammadnia}.

Finally, once $\cb^{\textmd{CS}}$ - the CS estimate of $\cb$ - has been obtained through any of the abovementioned schemes, we use (\ref{eq:freqsigmodel}) and (\ref{eq:vecXhat}) to obtain
\begin{eqnarray}\label{delta} \hat{\Xfd}&=&\underline{\hat{\Xfd}}-\Fm\cb^{\textmd{CS}}=\Xfd+[\Cfd-\Cfd^{\textmd{CS}}]+\Lambdam^{-1}\Zfd\nonumber\\
&\triangleq&\Xfd+\Efd^{\textmd{CS}}+\Lambdam^{-1}\Zfd\triangleq \Xfd+\Deltam,
\end{eqnarray}
where $\Efd^{\textmd{CS}} \triangleq \Cfd - \Cfd^{\textmd{CS}}$ and $\Deltam \triangleq \Efd^{\textmd{CS}}+\Lambdam^{-1}\Zfd$. $\hat{\Xfd}$ could be used to determine the data vector $\Xfd$ exactly, provided that no $\Delta(k)$ causes crossing of $\Xc(k)$ out of its ML decoding region (this issue will come up in Section \ref{dual}). Our subsequent objective is to scrutinize the general conditioning of the model itself by supplying our most reliable observations to the generic CS algorithm.

\section{Cherry Picking $\Omega_{m}$}\label{tone_selection_criteria}
An essential question now is how to select among the $2^{N}$ possible $\Omega_{m}$ (or $N \choose m$ if $m$ is fixed) in order to compute $\cb^{\textmd{CS}}$. In this connection we devise a \emph{reliability function} which associates a reliability estimate with each tone and thus lets us determine the $m$ most reliable tones to construct $\Omega_m$. A general strategy of CS techniques is to select these $m$ tones randomly for near-optimal performance \cite{Candes1}. Although possible in our scenario, such a strategy neglects the fact that our observations vary in their credibility and attest to whether or not they represent true frequency-domain measurements of $\Cfd$.\footnote{The measurements at tones $\Omega_m$ in (\ref{LASSO}) are used to determine $\cb$ and hence $\Cfd$. However, these measurements truly represent $\Cfd$ only if $\langle\hat{\Xc}(k)\rangle=\Xc(k)$ for $k\in\Omega_m$. We can not ascertain that it is true but we can calculate its probability.}

\begin{figure}[h!]%[!p]
\centering
\includegraphics[width=0.5\textwidth]{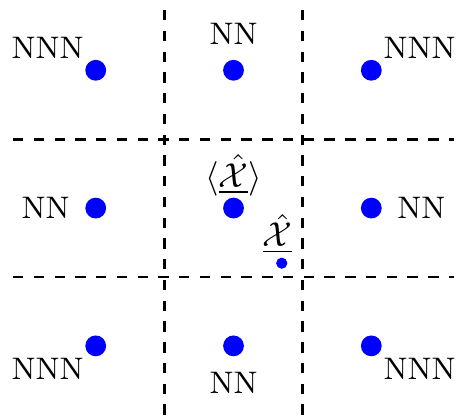}
\caption{An illustration of nearest neighbors (NN) and next nearest neighbors (NNN) of $\underline{\hat{\Xc}}$}
\label{fig:NNandNNN}
\end{figure}

\begin{figure}[h!]%[!p]
	\centering
		\includegraphics[width=0.5\textwidth]{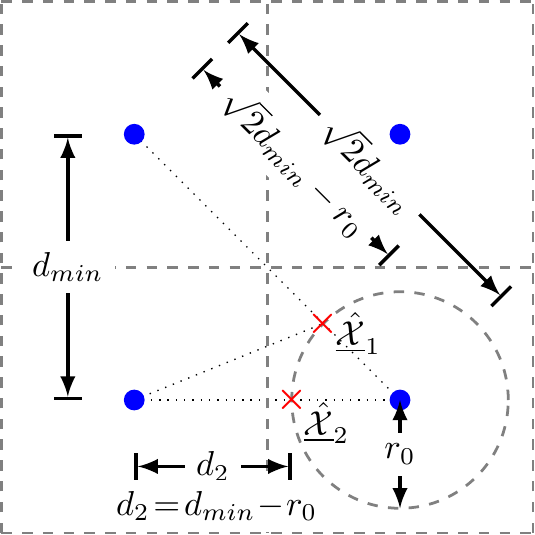}
	\caption{Reliability is not only a function of the magnitude but also the phase of the observation.}
	\label{fig:2reliabilities}
\end{figure}

Since we deal with each tone separately in what follows, we henceforth drop the $k$ index while preserving the italic notation to emphasize the scalar-wise operations in this section. With the receiver risking faulty decisions, it must devise a procedure to select the most reliable set of observations over which to sense. To this end consider the estimate $\underline{\hat{{\Xc}}}$ and the nearest constellation point $\langle\underline{\hat{{\Xc}}}\rangle$. The latter is in general surrounded by eight points which either belong to the set of nearest neighbors (NN) or the set of next nearest neighbors (NNN) as illustrated in Fig.~\ref{fig:NNandNNN}.
Now the selection of the most reliable set of observations could be done based on the relative posterior probability that $\Dfd$ equals
$\underline{\hat{\Xc}}-\langle\underline{\hat{\Xc}}\rangle$ to the probability that it equals some other difference vector
$\underline{\hat{\Xc}}-\Ac_{m}|\Ac_m\ne\langle\underline{\hat{\Xc}}\rangle$. For example, let $\mathfrak{R}=\frac{\Pr(\langle\underline{\hat{\Xc}}\rangle=\Xc)}{\Pr(\langle\underline{\hat{\Xc}}\rangle=\Ac_{\textmd{NN}})}$. The higher the value of $\mathfrak{R}$, the higher the reliability that $\Xc=\langle\hat{\Xc}\rangle$ as relative to the fact that $\hat{\Xc}$ should decode to $\Ac_{NN}$. From (\ref{eq:analog_estimate}), we can see that $\mathfrak{R}=\frac{\Pr(\Dc=\underline{\hat{\Xc}}-\langle\hat{\underline{\mathcal{X}}}\rangle)}{\Pr(\mathcal{D}=\hat{\underline{\mathcal{X}}}-\mathcal{A}_{\textmd{NN}})}$Now as mentioned earlier, we model $\Dc$ to be Gaussian circularly symmetric with variance $\sigma_{\Dc}^2$, then $f_\Dc(\Xc)=\frac{1}{\pi\sigma_\Dc^2}\exp(-\frac{1}{\sigma_\Dc^2} |\Xc|^2)$ and we can write, \begin{eqnarray*}\label{reliability_NN}
\mathfrak{R}\!&=&\!\exp{\left(\frac{-1}{\sigma_{D}^{2}}\left(|\underline{\hat{\Xc}}-\langle\underline{\hat{\Xc}}\rangle|^{2}-|\underline{\hat{\Xc}}-\Ac_{\textmd{NN}}|^{2}\right)\right)}
\end{eqnarray*}
Intuitively, the minimum certainty occurs at the boundary of the decision region and attains $\mathfrak{R}_{\min}=1$. At such tones, we would be highly skeptical of whether
$\Dc=\underline{\hat{\Xc}}-\langle\underline{\hat{\Xc}}\rangle$ or
$\Dc=\underline{\hat{\Xc}}-\Ac_{\textmd{NN}}$, and we would hence supply a plausibly false measurement to the CS algorithm. To avoid such unreliable measurements, assume we only choose the tones with respective perturbations $\underline{\hat{\Xc}}-\langle\underline{\hat{\Xc}}\rangle$ that are confined in the complex plane to a disk of radius $r_{o}$ (i.e., $|\hat{\underline{\Xc}}-\langle\underline{\hat{\Xc}}\rangle|\leq r_{o}$). In such a case, given the minimum distance between any two constellation points ($d_{\min}$), the minimum reliability would
increase to $\mathfrak{R}_{\min}=\frac{f_{\Dc}(r_o)}{f_{\Dc}(d_{\min}-r_{o})}$
in case the complex scalar $\underline{\hat{\Xc}}-\langle\underline{\hat{\Xc}}\rangle$ pointed in the direction of the nearest neighbor $\Ac_{\textmd{NN}}$, and to
$\mathfrak{R}=\frac{f_{\Dc}(r_o)}{f_{\mathcal{D}}\left(\sqrt{2}d_{\min}-r_{o}\right)}$
for the next nearest neighbor $\Ac_{\textmd{NNN}}$ when it points
in the direction of a decision region's corner. So while both $\hat{\Xc}_1$ and $\hat{\Xc}_2$ have the same distance $r_o$ from $\langle\hat{\Xc}_1\rangle=\langle\hat{\Xc}_2\rangle$, $\hat{\Xc}_1$ has a higher reliability than $\hat{\Xc}_2$ as it is farther from the nearest neighbor. Fig.~\ref{fig:2reliabilities} shows a part of constellation to illustrate the idea. This suggests a need to factor in the direction or \emph{phase} of the perturbation, $\theta_{\underline{\hat{\Xc}}-\langle\underline{\hat{\Xc}}\rangle}$, in assessing its reliability in addition to the magnitude-dependent pdf $f_{\Dc}$. Defined axiomatically, the reliability of a measurement
at each tone is then a function $\mathfrak{R}$ that maps a $3$-tuple
$(|\underline{\hat{\Xc}}-\langle\underline{\hat{\Xc}}\rangle|,\theta_{\hat{\underline{\Xc}}-\langle\underline{\hat{\Xc}}\rangle},\lambda^{-1})$ into $\mathbb{R}_{[1,\infty]}$ (i.e., a function of the magnitude of the observation, its phase and the channel gain at that tone).

Ultimately, we would choose our measurements according to the tones associated with the highest $m$ reliability outputs, i.e.,
\begin{eqnarray}\label{ordered_reliability}
\Omega_{m}\triangleq
\arg\left\{\mathfrak{R}_{i:N}\right\}_{i=N-m+1}^{N}
\end{eqnarray}
to sense over, where $\mathfrak{R}_{i:N}$ denotes the $i$th-order statistic in the vector $\bm{\mathfrak{R}}$ \cite{David}. With this selection of $\Omega_{m}$, the locations of the measurement tones correspond to the indices of the highest $m$-order statistics of $N$ random variables in $\bm{\mathfrak{R}}$. As mentioned previously, each of these variables is a function of the $3$-tuple above, and whereas the first two are uncorrelated across the tones, this does not generally hold for the third, i.e., $\lambda(k)$.

In fact, assuming $L_{\hb}$ channel taps with a uniform power-delay profile, then the absolute autocorrelation of the channel across tones $k$ and $l$ can be expressed as $|\mathbb{E}[\lambda(k)\lambda^{H}(l)]|=\textmd{sinc}\left(\pi L_{\hb}(k-l)/N\right)$ \cite{SVD}. Hence, only for sufficiently large $L_{\hb}$ can we assume that the channel gains are uncorrelated. Otherwise, the set of reliable tones $\Omega_{m}$ deviates from a uniformly random tone selection model typically assumed in the literature \cite{Candes1,Candes2}, and reliable tones would instead come in clusters corresponding to strong channel gains. The efficiency of CS in this case might be reduced\footnote{Nonetheless some methods such as FBMP are not much hindered by this fact \cite{Schniter2}.}.
\subsection{Criteria for Evaluating $\mathfrak{R}$}\label{intro_criteria}
Using the reasoning based on the scalar-wise likelihood ratio defined in (\ref{reliability_NN}), an exact expression for the reliability could be a direct generalization of (\ref{reliability_NN}), namely, \begin{eqnarray}\label{reliability_exact}
\mathfrak{R}^{\text{exact}}=\frac{f_{\Dc}(\underline{\hat{\Xfd}}-\langle\underline{\hat{\Xfd}}\rangle)}{\sum_{{m=0, \Ac_m\neq
\langle\underline{\hat{\Xc}}\rangle}}^{M-1}f_{\Dc}(\underline{\hat{\Xc}}-\Ac_{m})},
\end{eqnarray}
Unfortunately, this pursuit for exact reliability computation is inefficient since it requires $N\!M$ evaluations of $f_{\Dc}(\cdot)$, which grows with the constellation size $M$, even though the probability of a perturbation exceeding the first tier of eight surrounding neighbors (i.e., the nearest and next nearest neighbors illustrated in Fig. \ref{fig:NNandNNN}) is insignificant. As such, we can truncate the computation above to the first tier, denoting its result by $\mathfrak{R}^{\textmd{trunc}}$, with a minor effect on the performance.

Two simpler reliability functions are also worth mentioning. The first is solely based on the probability of the perturbation (i.e., $f_{\Dc}(|\hat{\underline{\Xc}}-\langle\underline{\hat{\Xc}}\rangle|)$) and is hence completely blind to the direction of the perturbation in the constellation's plane, while the other one intuitively takes this extra phase information into account by defining a square centered at $\langle\underline{\hat{\Xc}}\rangle$ as the reliable region, hence having the ability to favor perturbations with larger magnitudes if $\theta_{\underline{\hat{\Xc}}-\langle\underline{\hat{\Xc}}\rangle}$ were close to $\pi/4$, i.e., if they pointed to the next nearest neighbor. We denote these reliability functions by $\mathfrak{R}^{\circ}$ and $\mathfrak{R}^{\Box}$, respectively, motivated by the geometric shape they define. In the next section, a more rigorous approach is taken to justify when such simpler methods can be used.
\subsection{Analysis of Truncated and Approximate Reliability Criteria}
Dropping the tone index, assume that $\underline{\hat{\Xc}}$ is an observation that falls among four points in an $M$-QAM constellation such that $\underline{\hat{\Xc}}\in\mathbb{Q}$. Let $\underline{\hat{\Xc}}-\langle\underline{\hat{\Xc}}\rangle\triangleq r e^{\jmath\theta}$ be the polar representation of this point with the origin at $\langle\underline{\hat{\Xc}}\rangle$, such that $r\triangleq |\underline{\hat{\Xc}}-\langle\underline{\hat{\Xc}}\rangle|\in[0,\frac{1}{\sqrt{2}}d_{\min}]$ and $\theta\triangleq\theta_{\underline{\hat{\Xc}}-\langle\underline{\hat{\Xc}}\rangle}\in[0,\pi/2]$. We are interested in a more abstract expression of the truncated Bayesian reliability function, $\mathfrak{R}^{\textmd{trunc}}$, one that defines its output by only acting on $\underline{\hat{\Xc}}-\langle\underline{\hat{\Xc}}\rangle$ while taking the relative position in the constellation implicitly into account as well\footnote{This could be similarly carried out to the non-truncated function (\ref{reliability_exact}) albeit with an unnecessary inflation of expressions with hardly any additional insight.}.

\begin{figure}[h!]%[!p]

\centering
\includegraphics[width=0.5\textwidth]{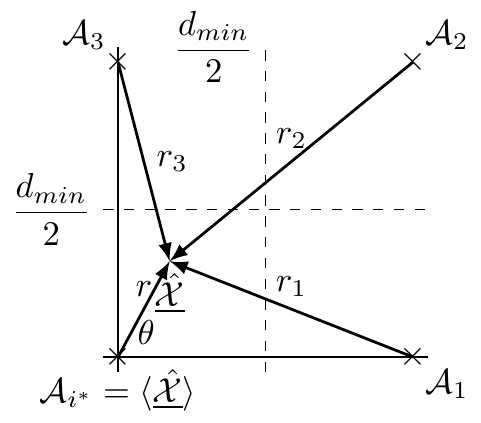}
\caption{Defining all $f_{\mathcal{D}}(r_{i})$ in terms of $r$ and $\theta$.}
\label{r_theta}
\end{figure}

By the definition of $r$, and by referring to Fig.~\ref{r_theta} which again shows a part of the constellation diagram (similar to Fig.~\ref{fig:2reliabilities}), the distances between $\underline{\hat{\Xc}}$ and the other three \emph{competing} constellation points are
\begin{eqnarray}
r_{1}(r,\theta)=\sqrt{r^{2}-2rd_{\min}\cos\theta+d_{\min}^{2}},\label{r1}
\end{eqnarray}
\begin{eqnarray}
r_{2}(r,\theta)=\sqrt{r^{2}-2rd_{\min}(\cos\theta+\sin\theta)+2d_{\min}^{2}},\label{r2}
\end{eqnarray}
and
\begin{eqnarray}
r_{3}(r,\theta)=\sqrt{r^{2}-2rd_{\min}\sin\theta+d_{\min}^{2}}.\label{r3}
\end{eqnarray}
\begin{figure}[h!]%[!p]

\centering
\includegraphics[width=0.5\textwidth]{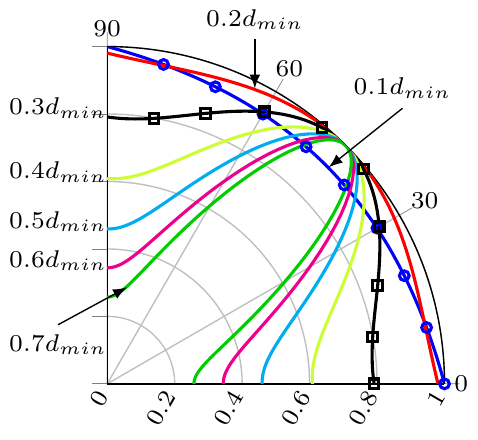}
\caption{$\mathfrak{R}^{\text{trunc}}(r,\theta)$ in (\ref{R_trunc}) normalized and plotted on the first quadrant for $\sigma_{\mathcal{D}}^{2}\!=\!0.2d_{\min}$ and evaluated at $r\!=0.1d_{\min},0.2d_{\min},$ $\ldots, 0.7d_{\min}$. Note that $\theta$ varies from $0^\circ$ to $90^\circ$.}
\label{Exact_G_function_fig}
\end{figure}
We neglect detailing their phases, $\theta_{1},\theta_{2},$ and
$\theta_{3}$, since they have no effect on the results, although $r_{1},r_{2}$, and $r_{3}$ are clearly functions of $r$ and $\theta$ as portrayed in Fig. \ref{r_theta}. In effect, $\mathfrak{R}^{\textmd{trunc}}$ reduces to
\begin{align}
\mathfrak{R}^{\textmd{trunc}}(r,\theta)=&f_{\Dc}(r)/\sum_{i=1}^{3}f_{\Dc}(r_{i}(r,\theta))\nonumber\\
=&e^{\frac{-r^{2}}{\sigma_{\Dc}^{2}}}/\Big[e^{\frac{-1}{\sigma_{\Dc}^{2}}(r^{2}-2rd_{\min}\cos\theta+d_{\min}^{2})}+\nonumber\\
&e^{\frac{-1}{\sigma_{\Dc}^{2}}\left(r^{2}-2rd_{\min}\left(\cos\theta+\sin\theta\right)+2d_{\min}^{2}\right)}+e^{\frac{-1}{\sigma_{\Dc}^{2}}\left(r^{2}-2rd_{\min}\sin\theta+d_{\min}^{2}\right)}\Big]\nonumber.
\end{align}
Canceling out the common function $f_{\Dc}(r)$ appearing in all terms above, and collecting common terms yields
\begin{eqnarray}\label{R_trunc}
\mathfrak{R}^{\textmd{trunc}}(r,\theta)\!\!\!&=&\!\!\!\Big(e^{-\frac{d_{\min}^{2}}{\sigma_{\Dc}^2}}[e^{\frac{1}{\sigma_{\Dc}^{2}}\left(2d_{\min}r\cos\theta\right)}+e^{\frac{1}{\sigma_{\Dc}^{2}}\left(2d_{\min}r\left(\cos\theta+\sin\theta\right)-d_{\min}^{2}\right)}+e^{\frac{1}{\sigma_{\Dc}^{2}}\left(2d_{\min}r\sin\theta\right)}]\Big)^{-1}\nonumber\\
&\triangleq&\!\!\!\!\left(\beta\left[\alpha^{\cos\theta}\!+\!\alpha^{\sin\theta}\!+\!\beta\alpha^{\cos\theta+\sin\theta}\right]\right)^{-1}
\end{eqnarray}
for the first quadrant, $\theta\in[0,\pi/2]$, in the complex plane, where $\alpha\triangleq\exp(2d_{\min}r/\sigma_{\Dc}^{2})$ and $\beta\triangleq\exp(-d_{\min}^{2}/\sigma_{\Dc}^{2})$. Although clearly a function of $r$, we will treat $\alpha$ as a constant (i.e., evaluated at a fixed magnitude, $r$) when we wish to focus on $\mathfrak{R}^{\textmd{trunc}}$ as an explicit function of $\theta$, say $g(\theta)$\footnote{The reliability on other quadrants is obtained by a basic reflection of $g(\theta)$ about the vertical and horizontal axes, or more simply by mapping $\theta\in[0,2\pi]$ back to $\theta\in[0,\pi/2]$.}. This function is symmetric about $\theta\!=\!\pi/4$ and exhibits quite complex behavior with $r$ and $\theta$ as indicated in Fig. \ref{Exact_G_function_fig}.

Most importantly, our concern will be whether $\mathfrak{R}^{\textmd{trunc}}(r,\theta)$ is convex or concave with respect to $\theta$ at different regions of $r$. This is because $\mathfrak{R}^{\textmd{trunc}}(\theta)$ undertakes a fundamental shift in behavior as $r$ varies from $0$ to $\frac{1}{\sqrt{2}} d_{\min}$, and its approximation by basic trigonometric functions and geometric objects such as squares and circles depends on whether it is convex or concave with respect to $\theta$. Notice first that when $r\ll\sigma_{\Dc}^{2}/2d_{\min}$, $\alpha\approx1$ and hence $\mathfrak{R}^{\text{trunc}}\approx\frac{1}{\beta[2+\beta]}$ becomes relatively isotropic (i.e., independent of $\theta$) and therefore akin to $\mathfrak{R}^{\circ}$. Referring to Fig. \ref{Exact_G_function_fig} for example, the polar curve of the normalized reliability function $\mathfrak{R}^{\textmd{trunc}}(\theta)|_{r=0.1d_{\min}}$, evaluated at the smallest magnitude $r=0.1d_{\min}$, confirms this observation. (Refer to the curve with blue circular markers in Fig.~\ref{Exact_G_function_fig}).

\begin{figure}[h!]%[!p]

\centering
\includegraphics[width=0.5\textwidth]{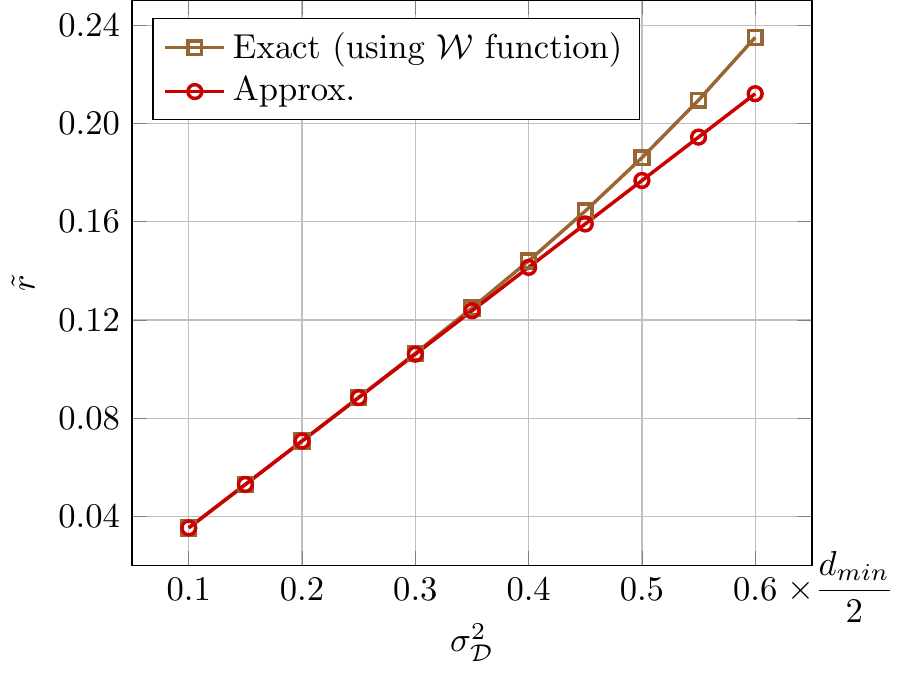}
\caption{Comparison of $\tilde{r}$ in (\ref{Lambert_roots}) and its approx. in (\ref{Lambert_approx}) as a function $\sigma_{\Dc}^{2}$ expressed as a ratio of $d_{\min}/2$ (at $d_{\min}=1$).}
\label{lambert_fig}
\end{figure}

In fact, as will be shown shortly, $\mathfrak{R}^{\textmd{trunc}}(r,\theta)$ will tend to even disfavor perturbations along $\pi/4$ (or $\pi/4+n\pi/2$, $n=1,\ldots,3$ in general) until it shifts gears and takes on its concave behavior with respect to $\theta$. For instance, the curve of the normalized function $\mathfrak{R}^{\textmd{trunc}}(\theta)|_{r=0.2d_{\min}}$ appearing in Fig.~\ref{Exact_G_function_fig} is actually convex (therefore assigning slightly higher reliability to perturbations in the direction of the next nearest neighbors having the same magnitude $0.2d_{\min}$), whereas the curves of the normalized function $\mathfrak{R}^{\textmd{trunc}}(\theta)|_{r=0.3d_{\min}}$ and beyond are concave. To pinpoint the location of this behavioral shift, we need to find
\begin{eqnarray}\label{partial_R}
\tilde{r}=\left\{r\in\mathbb{R}_{[0,\frac{\sqrt{2}}{2}d_{\min}]}:\frac{\partial^{2}\mathfrak{R}^{\textmd{trunc}}(r,\theta)}{\partial\theta^{2}}=0\right\},
\end{eqnarray}
where $\partial^{2}\mathfrak{R}^{\text{trunc}}(\theta)/\partial\theta^{2}$ is expressed as
\begin{eqnarray}
\frac{\partial^{2}\mathfrak{R}^{\textmd{trunc}}(\theta)}{\partial\theta^{2}}&=&\frac{1}{\beta}\Big(\frac{2\left(\beta\ln\alpha(\cos\theta-\sin\theta)\alpha^{\sin\theta+\cos\theta}-\ln\alpha\sin\theta\alpha^{\cos\theta}+\ln\alpha\cos\theta\alpha^{\sin\theta}\right)^{2}}{(\beta\alpha^{\sin\theta+\cos\theta}+\alpha^{\sin\theta}+\alpha^{\cos\theta})^{3}}\nonumber\\
&-&\big(\beta\ln^{2}\alpha(\cos\theta-\sin\theta)^{2}\alpha^{\sin\theta+\cos\theta}+\beta\ln\alpha(-\sin\theta-\cos\theta)\alpha^{\sin\theta+\cos\theta}\nonumber\\
&-&\ln\alpha\sin\theta\alpha^{\sin\theta}-\ln\alpha\cos\theta\alpha^{\cos\theta}+\ln^{2}\alpha\cos^{2}\theta\alpha^{\sin\theta}\nonumber\\
&+&\ln^{2}\alpha\sin^{2}\theta\alpha^{\cos\theta}\big)/\left(\beta\alpha^{\sin\theta+\cos\theta}+\alpha^{\sin\theta}+\alpha^{\cos\theta}\right)^{2}\Big)\label{detailed_partial}
\end{eqnarray}
Clearly, this is a daunting task. However, we can reduce it to finding the root, $\tilde{r}$, which satisfies (\ref{partial_R}) when evaluated at $\theta\!=\!\pi/4$, since our main concern is whether or not $\mathfrak{R}^{\textmd{trunc}}(r,\theta)$ will be tapered along $\theta=\pi/4$, and this will fortunately result in many cancelations in (\ref{detailed_partial}) due to symmetry about this point. Pursuing this reduces (\ref{partial_R}) to solving
\begin{eqnarray}
\sqrt{2}(\beta\alpha^{\frac{\sqrt{2}}{2}}+1)-\ln\alpha=0.
\end{eqnarray}
Expanding this into the original parameters implies that we need to find $\tilde{r}$ such that
\begin{eqnarray}\label{concavity_condition}
\frac{\sqrt{2}d_{\min}\tilde{r}}{\sigma_{\Dc}^{2}}-e^{\frac{\sqrt{2}d_{\min}\tilde{r}-d_{\min}^{2}}{\sigma_{\Dc}^{2}}}-1=0
\end{eqnarray}
which cannot be solved explicitly in terms of $\tilde{r}$. Rather, by means of a proper substitution, it can be put in the implicit form $g(\tilde{r})e^{g(\tilde{r})}\!=\!q$, where $q$ is independent of $\tilde{r}$ and expressed using the primary branch $\Wc_{0}$ of Lambert's $\Wc$-function \cite{Lambert}. The explicit solution to the previous form can be expressed as $g(\tilde{r})\!=\!\Wc_{0}(q)$, and the desired explicit expression $\tilde{r}$ is obtained by back-substitution (Refer to \ref{app2} for details). Ultimately, we can show that
\begin{eqnarray}\label{Lambert_roots}
\tilde{r}=\frac{-\sqrt{2}\sigma_{\Dc}^{2}}{2d_{\min}}\left(\Wc_{0}\left(-e^{1-\frac{d_{\min}^{2}}{\sigma_{\Dc}^{2}}}\right)-1\right).
\end{eqnarray}
Furthermore, as $\Wc_{0}(0)\!=\!0$ and $d_{\min}/2\!>\!\sigma_{\Dc}^{2}$, it is clear that $d_{\min}^{2}/\sigma_{\Dc}^{2}>2d_{\min}$, and that the argument of $\Wc_{0}$ quickly approaches zero from the left as $\sigma_{\Dc}^{2}$ diminishes, resulting in the following accurate approximation of (\ref{Lambert_roots}):
\begin{eqnarray}\label{Lambert_approx}
\tilde{r}\approx\frac{\sqrt{2}}{2}\frac{\sigma_{\Dc}^{2}}{d_{\min}}
\end{eqnarray}
for small $\sigma_{\Dc}^{2}$ relative to $d_{\min}/2$. Fig.~\ref{lambert_fig} plots (\ref{Lambert_roots}) and its approximation (\ref{Lambert_approx}) as functions of $\sigma_{\Dc}^{2}$. Using the approximation for simplicity, $\tilde{r}$ then splits the behavior of $\mathfrak{R}^{\textmd{trunc}}(r)|_{\theta=\pi/4}$ into two regions, supported by the intervals $\mathfrak{r}_{1}\approx\{r\in[0,\frac{\sqrt{2}}{2}\frac{\sigma_{\Dc}^{2}}{d_{\min}}]\}$ and $\mathfrak{r}_{2}\approx\{r\in[\frac{\sqrt{2}}{2}\frac{\sigma_{\Dc}^{2}}{d_{\min}},\frac{\sqrt{2}}{2}d_{\min}]\}$. This result explains why $\mathfrak{R}^{\textmd{trunc}}(r,\theta)$ in Fig.~\ref{Exact_G_function_fig} first resembles a circular shape akin to $\mathfrak{R}^{\circ}$ and then inflates along the diagonals deforming into a square-shaped reliability region, $\mathfrak{R}^{\Box}$, as can be seen when $r=0.3d_{\min}$ in Fig.~\ref{Exact_G_function_fig} (see the black curve with square-shaped markers). Subsequently, as the magnitude, $r$, of the perturbation increases and approaches the decision boundaries, $\mathfrak{R}^{\text{trunc}}(r,\theta)$ inflates outwards, resembling pointy leaves that can be modeled as $\mu+(1-\mu)\cos(4\theta+\pi)$, where $\mu\!\in[1/2,1]\!>\!1-\mu\!>\!0$. The analysis also provides restrictions for when square-like reliability regions suggested in the literature (such as \cite{Quasi}) can be justifiable.
\subsection{Dual-stage construction of $\Omega_{m}$}\label{dual}
The reader will notice that our primary objective so far in selecting $\Omega_{m}$ was based on minimizing the probability of incorrect measurements, i.e.,
\begin{align}\label{objective1}
\Omega_{m}=\arg\max_{\tilde{\Omega}_{m}}
\Pr\left(\tilde{\Omega}_{m}\subset\Omega_{T}\right).
\end{align}
This is no doubt a necessary choice to preserve the success of the recovery algorithm as a whole, although we know that a more generic criterion, that is, one that is not at risk of using incorrect observations, would seek the tones with the maximum
\textmd{CNR}\footnote{In other words, in a generic CS algorithm in which all measurements are 100\% reliable, the most effective measurements are the ones which maximize the CNR.}, i.e., (see (\ref{lossless_CS_model}))
\begin{eqnarray}\label{objective2}
\Omega_{m}&=&\arg\max_{\tilde{\Omega}_{m}}
\frac{\|\Psim_{\tilde{\Omega}_{m}}\cb\|_{2}^{2}}{\|\Zfd^\prime_{\tilde{\Omega}_{m}}\|_{2}^{2}}
=\arg\max_{\tilde{\Omega}_{m}} \sum_{k\in\tilde{\Omega}_{m}} \frac{|\Cc(k)|^{2}}{|\lambda^{-1}(k)\Zc(k)|^{2}}.
\end{eqnarray}
Obviously, there is a conflicting interest between (\ref{objective1}) and (\ref{objective2}), as the former frequently seeks smaller perturbations
$\underline{\hat{\Xc}}-\langle\underline{\hat{\Xc}}\rangle$, since they are the most likely to equal $\Dc$, while the latter seeks the largest perturbations to maximize the \textmd{CNR} for enhanced estimation performance.

This prompts us to consider a second recovery stage (i.e., another CS iteration) that uses (\ref{objective2}) and produces a \emph{new} subset of selected tones, denoted by $\Omega_{m}^{\textmd{CS}}$. The second recovery stage takes in $\Omega_m$ obtained from (\ref{objective1}) and uses it with any of the mentioned CS recovery algorithms to get a CS estimate of $\Cc$, denoted by $\Cc^{\textmd{CS}}$. This lets us achieve the corrected decoding decision $\langle\underline{\hat{\Xc}}-\Cc^{\textmd{CS}}\rangle$ allowing us to have a higher confidence that  $\Dc=\underline{\hat{\Xc}}-\langle\underline{\hat{\Xc}}\!-\Cc^{\textmd{CS}}\rangle$ compared to the primary assumption that $\Dc=\underline{\hat{\Xc}}-\langle \underline{\hat{\Xc}}\rangle$. This is because the decoding error in $\langle \underline{\hat{\Xc}}\rangle=\langle \Xc+\mathcal{D}\rangle=\langle\Xc+\Cc+\lambda^{-1}\Zc\rangle$ depends
on the value of $\Cc$, whereas the error in $\langle \underline{\hat{\Xfd}}-\Cc^{\textmd{CS}}\rangle=\langle \Xc+\Delta\rangle=\langle\Xc+\Ec^{\textmd{CS}}+\lambda^{-1}\Zc\rangle$ depends on the \emph{estimation error} $\Ec^{\textmd{CS}}=\Cc-\Cc^{\textmd{CS}}$ of $\Cc$ which is expected to be smaller than $\Cc$ itself. These results follow from (\ref{X_bar_estimate}) and (\ref{delta}).

As illustrated in Figs.~\ref{fig:CNR_in_highSNR} and \ref{fig:CNR_in_lowSNR}, it is possible now to use these carriers that have the largest values of the perturbations, $\underline{\hat{\Xc}}-\langle\underline{\hat{\Xc}}-\Cc^{\textmd{CS}}\rangle$ (or even the carriers with the largest values of $\Cc^{\textmd{CS}}$), as the new CS measurements, without worrying much
about how close $\underline{\hat{\Xc}}$ is to the decision boundaries. Note, however, that we never have access to $\Cc$, $\Zc$, or $\Ec^{\textmd{CS}}$, and therefore we can rely only on observable variables, such as $\Dc$ and $\Delta$, to practically maximize the $\textmd{CNR}$. More importantly, these variables themselves are not always obtainable, since it is not necessarily the case that $\Dc=\underline{\hat{\Xc}}-\langle\underline{\hat{\Xc}}-\Cc^{\textmd{CS}}\rangle$ or that $\Delta=\underline{\hat{\Xc}}-\Cc^{\textmd{CS}}-\langle\underline{\hat{\Xc}}-\Cc^{\textmd{CS}}\rangle$ (which is the main reason that we repeat CS over a subset of measurements). Instead, we have to rely on the \emph{observable} variables, $\underline{\hat{\Xc}}-\langle\underline{\hat{\Xc}}-\Cc^{\textmd{CS}}\rangle$ and $\Cc^{\textmd{CS}}$, for this task. The availability of  these two observable variables grants us the flexibility of computing \textmd{CNR} in two different ways, both of which are suitable for different scenarios. The first is suited for a high SNR and low CS estimation quality (see Fig.~\ref{fig:CNR_in_highSNR}), while the second is suited for a low SNR and high CS estimation quality (see Fig.~\ref{fig:CNR_in_lowSNR}). Dropping the coefficient index, $k$, these are:
\begin{enumerate}

\item High SNR, large $\Ec^{cs}$:
\begin{eqnarray}
\widehat{\textmd{CNR}}^{\lambda}&=&\frac{|\,\underline{\hat{\Xc}}-\langle
\underline{\hat{\Xc}}-\Cc^{\textmd{CS}}\rangle|^{2}}{|\,\underline{\hat{\Xc}}-\Cc^{\textmd{CS}}-\langle
\underline{\hat{\Xc}}-\Cc^{\textmd{CS}}\rangle|^{2}}\nonumber\\
&=&\frac{|\mathcal{D}|^{2}}{|\Delta|^{2}}, \quad \textmd{iff} \:\:\langle
\underline{\hat{\Xfd}}-\Cc^{\textmd{CS}}\rangle=\Xc\\ &=&\frac{|\Cc+\lambda^{-1}\Zc|^{2}}{|\Ec^{\textmd{CS}}+\lambda^{-1}\Zc|^{2}}\quad\underset{\lambda^{-1}\rightarrow 0}{\asymp}\quad\frac{|\Cc|^{2}}{|\Ec^{\textmd{CS}}|^{2}}\nonumber
\end{eqnarray}
where the symbol $\asymp$ means that the expression on the left hand side of this symbol is asymptotically equal to the expression on the right hand side.
\begin{figure}[h!]%[!p]

\centering
\includegraphics[width=0.5\textwidth]{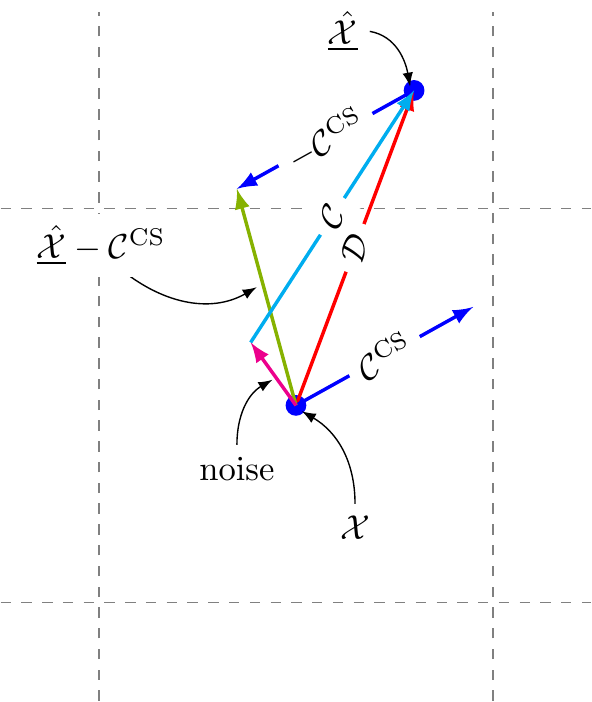}
\caption{Computing CNR for the case of high SNR and low CS estimation quality}
\label{fig:CNR_in_highSNR}
\end{figure}

\item Low SNR, small $\Ec^{cs}$:
\begin{eqnarray}
\widehat{\textmd{CNR}}^{\Ec^{\textmd{CS}}}&=&\frac{|\,\Cc^{\textmd{CS}}|^{2}}{|\,\hat{\underline{\Xc}}-\Cc^{\textmd{CS}}-\langle\underline{\hat{\Xc}}-\Cc^{\textmd{CS}}\rangle|^{2}}\nonumber\\
&=&\frac{|\Cc^{\textmd{CS}}|^{2}}{|\Delta|^{2}}, \quad \textmd{iff} \:\:\langle
\underline{\hat{\Xc}}-\Cc^{\textmd{CS}}\rangle=\Xc\\
\nonumber&=&\frac{|\Cc+\Ec^{\textmd{CS}}|^{2}}{|\lambda^{-1}\Zc+\Ec^{\textmd{CS}}|^{2}}\quad\underset{\Ec^{\textmd{CS}}\rightarrow 0}{\asymp}\quad\frac{|\Cc|^{2}}{|\lambda^{-1}\Zc|^{2}}.
\end{eqnarray}
\begin{figure}[h!]%[!p]

\centering
\includegraphics[width=0.5\textwidth]{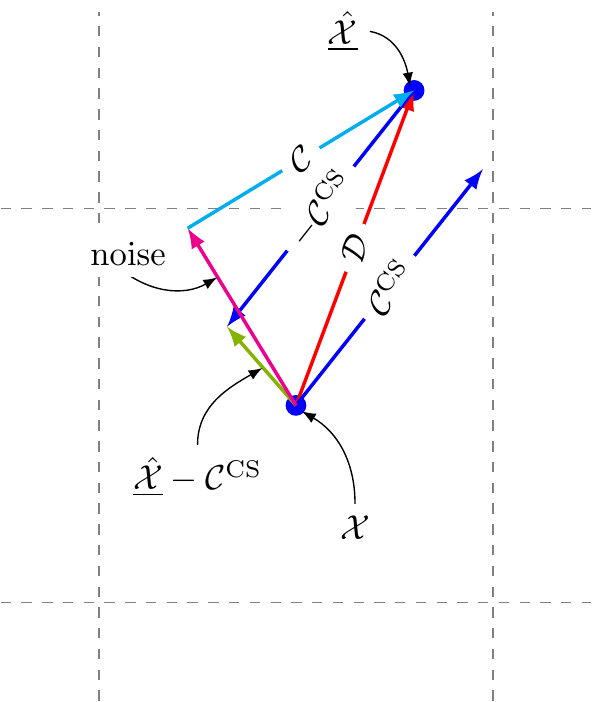}
\caption{Computing CNR for the case of low SNR and high CS estimation quality}
\label{fig:CNR_in_lowSNR}
\end{figure}
\end{enumerate}

Although the second ratio $\widehat{\textmd{CNR}}^{\Ec^{\textmd{CS}}}$ more vividly resembles the $\textmd{CNR}$ defined in (\ref{objective2}), the first ratio is more effective in this work as the inherent complexity of CS based methods justifies itself in severe clipping scenarios and hence expectedly higher CS error (i.e., large $\Ec^{\textmd{CS}}$). Consequently, we select the differential measurements corresponding to the maximum $|\Omega_{m}^{\textmd{CS}}|$ ratios\footnote{Obviously, the number of tones $|\Omega_{m}^{\textmd{CS}}|$ need not be equivalent to the original number
$|\Omega_{m}|$, and a wealth of possibilities emerges in relating these two parameters for optimal performance.},
\begin{eqnarray}
\Omega_{m}^{\textmd{CS}}=\arg\left\{\widehat{\textmd{\textbf{CNR}}}^{\lambda}_{i}\right\}_{i=N-|\Omega_{m}^{\textmd{CS}}|+1}^{N},
\end{eqnarray}
and the new CS model is
\begin{eqnarray}\label{new_CS_model}
\Yfd^\prime_{\Omega_{m}^{\textmd{CS}}}&=&\Sm_{\Omega_{m}^{\textmd{CS}}}(\underline{\hat{\Xfd}}-\langle\underline{\hat{\Xfd}}-\Cfd^{\textmd{CS}}\rangle)+\Zfd_{\Omega_{m}^{\textmd{CS}}}^\prime\nonumber \\
&=&\Psim_{\Omega_{m}^{\textmd{CS}}} \cb+\Zfd_{\Omega_{m}^{\textmd{CS}}}^\prime,
\end{eqnarray}
which produces an improved estimate, $\Cfd^{\textmd{rev}}$. This new estimate of the clipping distortion from a different subset of reliable and stronger measurements can then be subtracted from $\underline{\hat{\Xfd}}$, and a revised set of decoding decisions,
$\langle\underline{\hat{\Xfd}}-\Cfd^{\textmd{rev}}\rangle$, can be made.
\subsection{On selecting the cardinality $|\Omega_{m}|$}\label{cardinality}
\begin{figure}[h!]%[!p]
\centering
\includegraphics[width=0.5\textwidth]{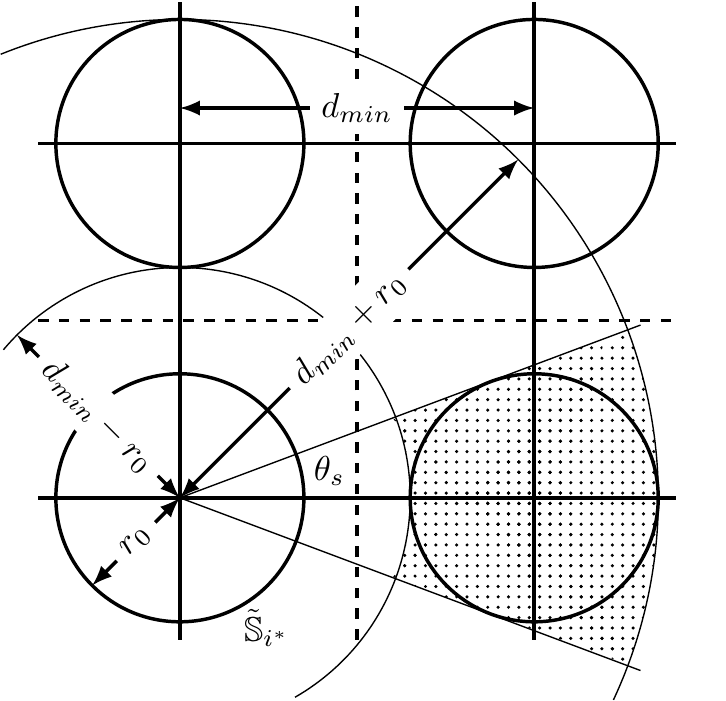}
\caption{Illustrating how the integral over the shaded region upper-bounds the integral over $\tilde{\mathbb{S}}_{i}$.}
\label{4circles_fig}
\end{figure}
In the method we proposed, it is assumed throughout the above discussion that the observations supplied to the CS algorithm were true measurements of the actual perturbations caused by clipping and additive noise. Although this is a mild restriction in practice, a guarantee must nonetheless be established
that any selected $\Omega_{m}$ and $|\Omega_{m}|$ according to Section \ref{tone_selection_criteria} will not result in CS failure. This requires that the number of measurements $|\Omega_{m}|$ be
both \emph{small} enough to minimize the probability of incorrect measurements, and also \emph{large}
enough relative to the sparsity level of $\cb$ to ensure meeting recovery bounds of CS.

To this end, we derive a simple lower bound on the probability of $\Omega_{m}\!\subset\!\Omega_{T}$ by deriving a lower bound on the probability that $\Dc$ is indeed equal to $\underline{\hat{\Xc}}-\langle\underline{\hat{\Xc}}\rangle$. We study this for two cases: when $\underline{\hat{\Xc}}-\langle\underline{\hat{\Xc}}\rangle$ is observed within a disk of radius $r_{o}$ from $\langle\hat{\underline{\mathcal{X}}}\rangle$, i.e., $\Omega_{m}\!=\!\{k\!\in\!\Omega\!:\!|\underline{\hat{\Xc}}(k)\!-\!\langle\underline{\hat{\Xc}}(k)\rangle|\!<\!r_{o}\}$, and when it is observed within a square of side length $2r_{o}$ centered at the QAM symbols $\Afd$. We focus on the first case since it is more difficult to estimate, and directly give the result of the second since it is comparably straightforward. To this end, define a \emph{safe} region, $\tilde{\mathbb{S}}_{i}=\{\Ac_{i}+\Uc\in \mathbb{C}:|\Uc|<r_{o}\}$, for decoding $\underline{\hat{\Xc}}(k)$ within its square \textmd{ML} decision region, $\mathbb{Q}_{i}$, and denote the collection of all these safe regions by $\mathbb{S}\!=\!\bigcup_{i=0}^{M-1}\tilde{\mathbb{S}}_{i}$. Our objective is to select $|\Omega_{m}|$ such that $\Pr\left(\Omega_{m}\subset\Omega_{T}\big||\Omega_{m}|\right)$ is high given a minimum amount of required measurement for CS success.

Dropping the tone index, this will require finding $\Pr(\Dc\in\mathbb{S})$, which requires evaluating an integral of $f_{\Dc}$ over (non-centered) discrete disks in the complex plane (see Fig.~\ref{4circles_fig}). Since this is difficult, we will use an upper bound based on evaluating $f_{\Dc}$ over centered disks that cover these regions and then slice out the irrelevant regions. More specifically, the integral $\int_{\Dc\in\tilde{\mathbb{S}}_{i}}$ over the disk $\tilde{\mathbb{S}}_{i}$ of a nearest neighbor will be bounded by the integral over the area highlighted by the shaded region in Fig.~\ref{4circles_fig}. This region  covers the difference between the outer and inner sectors defined by radii $d_{\min}+r_{o}$ and $d_{\min}-r_{o}$, respectively, and a common angle, $\theta_{s}=2\sin^{-1}\left(\frac{r_{o}}{d_{\min}}\right)$. In effect, the area of $\tilde{\mathbb{S}}_{i}$ is strictly less than $\frac{\theta_{s}}{2\pi}[\pi(d_{\min}+r_{o})^{2}-\pi(d_{\min}-r_{o})^{2}]$ and therefore $\Pr(\Dc\in\tilde{\mathbb{S}}_{i})<\frac{\theta_{s}}{2\pi}\left(F_{\Dc}(d_{\min}+r_{o})-F_{\Dc}(d_{\min}-r_{o})\right)$ for the nearest neighbor. Consequently,
\begin{eqnarray}
\Pr(\Omega_{m}\!\!\subset\!\Omega_{T}\big||\Omega_{m}|)\!\!\!\!&=&\!\!\!\!\Pr\left(\bigcap_{i=1}^{|\Omega_{m}|}\Omega_{m}(i)\in\Omega_{T}\right)\nonumber\\
\!\!\!\!&=&\!\!\!\!\Pr\left(\langle\underline{\hat{\Xc}}\rangle=\Xc\,\big||\,\underline{\hat{\Xc}}\!-\!\langle\underline{\hat{\Xc}}\rangle|\!<r_{o}\right)^{|\Omega_{m}|}=\Pr(\Dc=\underline{\hat{\Xc}}-\langle\underline{\hat{\Xc}}\rangle\big|\Dc\in\mathbb{S})^{|\Omega_{m}|}\nonumber\\
\!\!\!\!&=&\!\!\!\!\left(\frac{\Pr(|\Dc|<r_{o},\Dc\in\mathbb{S})}{\Pr(\Dc\in\mathbb{S})}\right)^{|\Omega_{m}|}=\left(\frac{\Pr(|\Dc|<r_{o})}{\Pr(\Dc\in\mathbb{S})}\right)^{|\Omega_{m}|}\label{ratio_prob}\\
&>&\!\!\!\!\mathbb{F}^{|\Omega_{m}|}_{|\Dc|}(r_{o})/\Big(\mathbb{F}_{|\Dc|}(r_{o})+\frac{8}{\pi}\sin^{-1}\!\frac{r_{o}}{d_{\min}}\left[\mathbb{F}_{|\Dc|}(d_{\min}\!+\!r_{o})\!-\!\mathbb{F}_{|\Dc|}(d_{\min}\!-\!r_{o})\right]\!\Big)^{\!|\Omega_{m}|}\nonumber\\
&>&\!\!\!\big(1-e^{\frac{-r_{o}^{2}}{2\sigma_{\Dc}^{2}}}\big)^{|\Omega_{m}|}/\Big(1-e^{\frac{-r_{o}^{2}}{2\sigma_{\Dc}^{2}}}\!+\!\frac{8}{\pi}\sin^{-1}\!\!\frac{r_{o}}{d_{\min}}\sinh\!\Big(\frac{r_{o}d_{\min}}{\sigma_{\Dc}^{2}}\Big)e^{\frac{-(d_{\min}^{2}+r_{o}^{2})}{2\sigma_{\Dc}^{2}}}\!\!\Big)^{|\Omega_{m}|}.
\end{eqnarray}
In the case where the average distortion is large and square regions $\mathbb{S}^{\Box}\!=\!\bigcup_{i=0}^{M-1}\tilde{\mathbb{S}}^{\Box}_{i}$ of side-length $2r_{o}$ are used, pursuing the same logic above we just replace the ratio in (\ref{ratio_prob}) with
\begin{eqnarray}
\frac{\Pr(\Dc\in{\tilde{\mathbb{S}}}^{\Box}_{i^{*}})}{\Pr(\Dc\in\mathbb{S}^{\Box})}\approx\frac{\left[1-2Q(\frac{r_{o}}{\sigma_{\Dc}})\right]^{2}}{\left[1-2Q(\frac{r_{o}}{\sigma_{\Dc}})\right]^{2}\!+\!4\left[Q(\frac{d_{\min}-r_{o}}{\sigma_{\Dc}})-Q(\frac{d_{\min}+r_{o}}{\sigma_{\Dc}})\right]^{2}}
\end{eqnarray}
and raise it to the power $|\Omega_{m}|$ to obtain $\Pr(\Omega_{m}\!\!\subset\!\Omega_{T}\big||\Omega_{m}|)$, where $Q$ is the familiar tail probability function. The user must then choose $|\Omega_{m}^{\,\tau}|=\arg\max_{|\Omega_{m}|} [\Pr(\Omega_{m}\subset\Omega_{T}\big||\Omega_{m}|)>\tau]$ where $\tau$ is selected so as to supply as much information to the CS algorithm as possible while remaining in a safe region of correct measurements. Furthermore, given a clipping threshold, $\gamma$, we have an expected sparsity level, $\mathbb{E}[|\Ic_{\cb}|]$, and variance $\sigma_{|\mathcal{I}_{\cb}|}^{2}$, which need to be taken into account when using sparse recovery techniques \cite{Candes1}. We will denote this minimum required number of frequency observations to recover an $|\Ic_{\cb}|$-sparse vector in time by $|\Omega_{m}^{\gamma}|$ to stress its strong dependence on $\gamma$, and take $|\Omega_{m}|=\max(|\Omega_{m}^{\tau}|,|\Omega_{m}^{\gamma}|)$. The same can be done with the optional second stage $|\Omega_{m}^{\textmd{CS}}|$.

Suppose however, after taking all the protective measures thus far, that an incorrect measurement was nonetheless supplied to the CS algorithm. Does this result in CS failure? Luckily, in this application the answer is no. Recall first the decoding-error vector $\Efd\triangleq\Xfd-\langle\underline{\hat{\Xfd}}\rangle$ used to motivate (\ref{difference_equivalence}). When a decoding error is made at the $k$th coefficient, an incorrect measurement $\underline{\hat{\Xc}}(k)-\langle\underline{\hat{\Xc}}(k)\rangle=\Xc(k)+\Dc(k)-\langle\underline{\hat{\Xc}}(k)\rangle\triangleq\Dc(k)+\Ec(k)$ is supplied and it follows from (\ref{difference_equivalence}) that the incorrect measurement has no impact on the performance of CS. Note that the nonzero entries of $\Efd$ are \emph{quantized} and \emph{bounded} since $\Xfd$ and $\langle\underline{\hat{\Xfd}}\rangle\in\Afd^{N}$. Furthermore, assuming most equalized coefficients $\underline{\hat{\Xc}}(k)$ are decoded correctly, $\Efd$ is also sparse.

The general differential model (i.e., the one that is not confined to the carriers in $\Omega_{m}$) becomes
\begin{eqnarray}
\underline{\hat{\Xfd}}-\langle\underline{\hat{\Xfd}}\rangle=\Cfd+\Lambdam^{-1}\Zfd+\Efd \label{structured_noise_model}
\end{eqnarray}
The first term added to $\Cfd$ is the (dense) Gaussian noise vector, while the second is a \emph{structured}-noise term. Equation (\ref{structured_noise_model}) matches the model in \cite{structured_noise2}, and it is demonstrated therein how it is possible to recover $\Cfd$ from such noise via variations in the CS algorithm. The main results of the paper are summarized in the algorithm listed in Table \ref{tab:ourmethod}.

\begin{table}[h!]%[!p]
\centering
\small\addtolength{\tabcolsep}{-8pt}
\begin{tabular}{l}
  \hline
  \textbf{parameters:} $N$, $M$, $\gamma$, $\mu$, $\tau$, $\sigma_{z}^{2}$, $\epsilon$\\
  \textbf{input:} $\Yfd$, $\Lambdam$,\\
  \textbf{output:} $\hat{\Xfd}=\langle\underline{\hat{\Xfd}}-\Cfd^{\textmd{CS}}\rangle$\\
  1. Compute $\underline{\hat{\Xfd}}$, $\langle\underline{\hat{\Xfd}}\rangle$, $|\underline{\hat{\Xfd}}-\langle\underline{\hat{\Xfd}}\rangle|$, $\theta_{\underline{\hat{\bm{\Xc}}}-\langle\underline{\hat{\bm{\Xc}}}\rangle}$, $\sigma_{\bm{\Dc}}^{2}$, $|\Omega_{m}^{\tau}|$, $|\Omega_{m}^{\gamma}|$\\
  2. Case: Severe clipping, for $k=1,\ldots,N$\\ \indent $\mathfrak{R}(k)\rightarrow\left(\beta(\alpha^{\cos\theta}+\alpha^{\sin\theta}+\beta\alpha^{\cos\theta+\sin\theta})\right)^{-1}$; $\alpha=e^{\frac{2d_{\min}r}{\sigma_{\Dc(k)}^{2}}}$, $\beta=e^{\frac{-d_{\min}}{\sigma_{\Dc(k)}^{2}}}$\\
  3. Case: Mild clipping, for $k=1,\ldots,N$,\\ \indent if $|\underline{\hat{\Xc}}(k)-\langle\underline{\hat{\Xc}}(k)\rangle|<\frac{\sqrt{2}}{2}\frac{\sigma_{\Dc(k)}^{2}}{d_{\min}}$,
  $\mathfrak{R}(k)\rightarrow f_{\Dc(k)}(\underline{\hat{\Xc}}(k)-\langle\underline{\hat{\Xc}}(k)\rangle)$\\
  \indent else $\mathfrak{R}(k)\rightarrow$ $f_{\Dc(k)}(\underline{\hat{\Xc}}(k)-\langle\underline{\hat{\Xc}}(k)\rangle)\cdot\left[\mu+(1-\mu)\cos(4\theta_{\underline{\hat{\Xc}}(k)-\langle\underline{\hat{\Xc}}(k)\rangle}+\pi)\right]$\\
  4. $|\Omega_{m}|=\max(|\Omega_{m}^{\tau}|,|\Omega_{m}^{\gamma}|)$\\
  5. Use any sparse recovery method (e.g. WPAL (\ref{LASSO}) or FBMP \!\cite{Schniter2})\\
   \indent on $\underline{\hat{\Xfd}}-\langle\underline{\hat{\Xfd}}\rangle$ over $\Omega_{m}$ and decode, i.e., $\hat{\Xfd}=\langle\underline{\hat{\Xfd}}-\Cfd^{\textmd{CS}}\rangle$ \\
  6. (Optional): \\
\hspace{18pt}a. $\textmd{CNR}^{\textmd{rev}}=\frac{|\underline{\hat{\Xfd}}-\langle
\underline{\hat{\Xfd}}-\Cfd^{\textmd{CS}}\rangle|}{|\underline{\hat{\Xfd}}-\Cfd^{\textmd{CS}}-\langle
\underline{\hat{\Xfd}}-\Cfd^{\text{CS}}\rangle|}$,\\
\hspace{18pt}b. Select $\Omega_{m}^{\textmd{CS}}$ using $\textmd{CNR}^{\text{rev}}$\\
\hspace{18pt}c. Perform CS on $\underline{\hat{\Xfd}}-\langle\underline{\hat{\Xfd}}-\Cfd^{\textmd{CS}}\rangle$ over $\Omega_{m}^{\textmd{CS}}$\\
\hline
\end{tabular}
	\caption{The proposed method}
	\label{tab:ourmethod}
\end{table}
\section{Simulation Results}\label{sim_results}
In this section, we perform several different experiments to show the effectiveness of the proposed technique. The methods proposed in this paper were tested on an $256$ subcarrier OFDM signal, modulated using $64$-QAM. The signal was subject to a block-fading, frequency-selective, Rayleigh channel
model, subject to varying noise and clipping ratios (CR) defined as CR$=\gamma/\sigma_{x}$ \cite{lin2005clipping}. Here, $\sigma_x$ is the standard deviation of the OFDM signal. Available
packages for convex programming \cite{Boyd} and Fast Bayesian Matching Pursuit (FBMP)
\cite{Schniter2} were used to implement the sparse-recovery algorithms. The undistorted phase property (as utilized in \cite{Safadi2}) is utilized while implementing FBMP and hence the modified version is termed phase augmented-FBMP (PAFBMP). Lastly, we refer to the \emph{second stage} defined in \ref{dual} as a \emph{corrective} action on the first estimation operation, and label its output by C-WPAL or C-PAFBMP for the cases when WPAL \cite{Safadi2} and PAFBMP \cite{Schniter2}
are used, respectively.

\begin{figure}[h!]%[!p]

\centering
\includegraphics[width=0.7\textwidth]{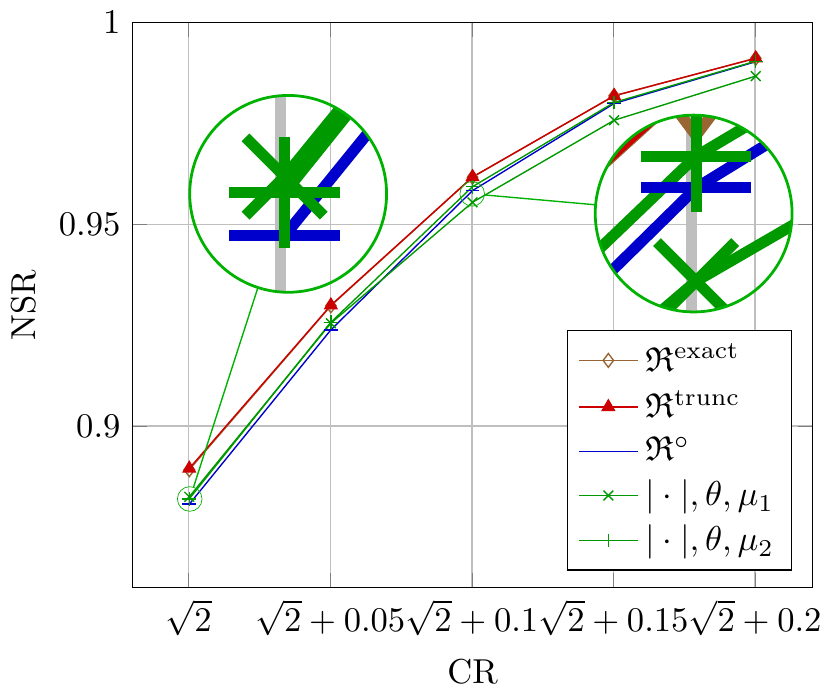}
\caption{NSR vs. CR for the various reliability
functions defined in Section \ref{intro_criteria}. $E_b/N_0=20$dB, $|\Omega_m|=64$, $\mu_{1}=0.65$, $\mu_{2}=0.95$.}
\label{Reliability_fig}
\end{figure}

\begin{figure}[h!]%[!p]

\centering
\includegraphics[width=0.7\textwidth]{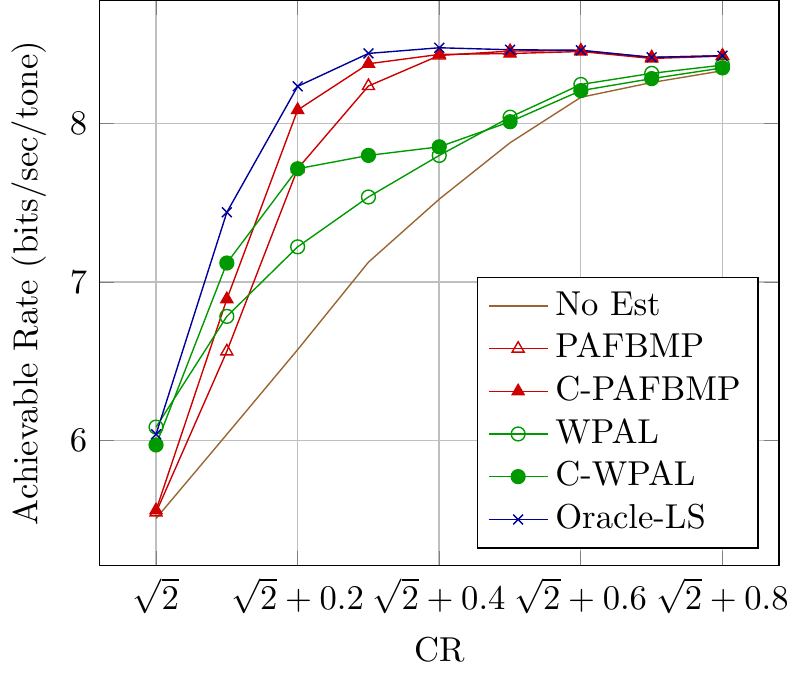}
\caption{Achievable Rate as a function of the clipping ratio CR$=\gamma/\sigma_{x}$.}
\label{capacity_vs_gamma}
\end{figure}
\begin{figure}[h!]%[!p]

\centering
\includegraphics[width=0.7\textwidth]{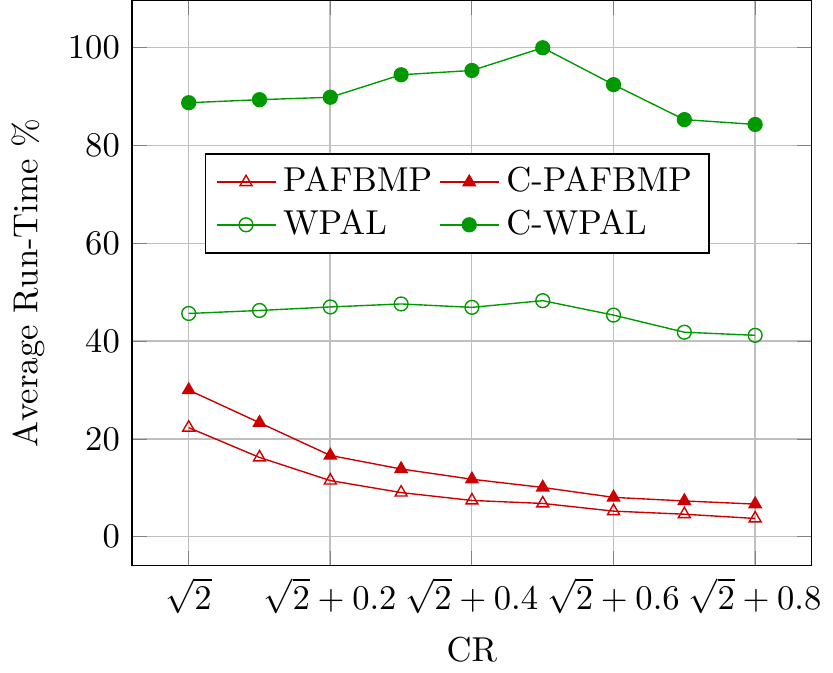}
\caption{Relative run time (max. = 100\%) at different clipping levels. $m_{\textmd{PAFBMP}}=0.25N$, $m_{\textmd{C-PAFBMP}}=0.39N$, $m_{\textmd{WPAL}}=0.25N$, $m_{\textmd{C-WPAL}}=0.39N$, $B=5$.}
\label{run_time}
\end{figure}
\subsection{Comparison of Reliability Criterias}\label{Capacity_sim}
In the first experiment, the reliability criterias proposed in this work are compared, including $\mathfrak{R}^{\text{exact}}$, $\mathfrak{R}^{\text{trunc}}$, $\mathfrak{R}^{\circ}$ and $\mathfrak{R}^{\Box}$. As a performance metric, we use normalized success rate (NSR) defined as $|\{k:\langle\hat\Xc(k)\rangle=\Xc(k),k\in\Omega_m\}|/|\Omega_m|$. The NSR depicts that among the $|\Omega_m|$ tones favoured by a particular reliability criteria, how many were actually within their corresponding correct decision regions. Fig. \ref{Reliability_fig} shows the result of this experiment. The results were plotted against a varying CR while $64$ most reliable carriers were sought keeping $E_b/N_0$ fixed at $20$dB. It is expected that as CR is increased, all reliability criterias will tend to improve. The simulation results confirm this intuition and also confirm the conjecture that the truncated reliability computation comes at little cost compared to
using the exact reliability function (\ref{reliability_exact}). Furthermore, it is shown for the parameters used for penalizing the circular
reliability function, $\mathfrak{R}^{\circ}$, by $\mu_{2}=0.95$ (which results in a square-like function such as the curve plotted in Fig. \ref{Exact_G_function_fig} for $r=0.3d_{\min}$) was more effective
than with a smaller value of $\mu_{1}=0.65$, for milder CR levels. On the other hand, with severe CR, smaller values of $\mu$ were better. This observation is highlighted by showing enlarged version of the graph for severe and milder CRs.

\subsection{Achievable Rate}
In this experiment, the ultimate performance measure considered was the achievable rate with and without the proposed method. Assuming ergodicity over the subcarriers, this rate can be expressed as $\frac{1}{N}\left(\sum_{i=0}^{N-1}\log 2 (1+|\lambda_i|^2\sigma_\Xc^2/(|\lambda_i|^2)\sigma_\Cc^2+\sigma_\Zc^2)\right)$ for the un-mitigated case, and as
$\frac{1}{N}\left(\sum_{i=0}^{N-1}\log 2 (1+|\lambda_i|^2\sigma_\Xc^2/(|\lambda_i|^2)\sigma_{\Cc-\hat\Cc}^2+\sigma_\Zc^2)\right)$, for the case when an estimate $\bm{\hat{\mathcal{C}}}$ of $\bm{\mathcal{C}}$ is obtained by an arbitrary method  \cite{Tellado_Book}. As a benchmark Oracle-LS is utilized, where the support of clipping signal is perfectly known and the active elements are estimated using LS estimate at the receiver. By using the WPAL (\ref{LASSO})\cite{Safadi2} and PAFBMP \cite{Schniter2} sparse recovery algorithms over
the exact reliability function $\mathfrak{R}^{\text{exact}}$, we compared the clipping mitigation results. The results in Fig. \ref{capacity_vs_gamma} show the superior performance of C-PAFBMP \cite{Schniter2} using the most reliable $39\%$
of carriers and the ability of C-WPAL to beat
all the techniques at CRs around $1.5$ using the most reliable $39\%$ carriers.
\subsection{Complexity}\label{complexity}
A practical comparison of relative run times for the tested algorithms is reported in Fig. \ref{run_time}. All times are scaled and represented as percentage of the maximum time required for recovery. We observe that PAFBMP has the least complexity among the compared schemes and this complexity reduces with increased CR.
\section{Conclusion}\label{conclusion}
A novel method for nonlinear distortion mitigation using pilotless sparse recovery techniques
was proposed. The method exploits the sparsity of the distortion in time domain to fully recover the signal without being influenced by incorrect ML-decoding decisions. The method adaptively senses over reliable subsets of observations of the distortion in frequency domain to perform the recovery. A new method of computing the reliability of each observation independently of the other $M-1$ candidates within a constellation was also proposed and tested. Through simulations, it is verified that the proposed scheme provides favourable results for clipping signal recovery and achieves a rate close to Oracle-LS based recovery.
\appendix
\section{Deriving $f_{\Dc}$}\label{app1}
Beginning with the fact that
\begin{eqnarray}
\sigma_{\Dc}^{2}=\sigma_{\Cc}^{2}+\lambda^{-\herm}\lambda^{-1}\sigma_{\Zc}^{2},\label{sigma_D}
\end{eqnarray}
can be made, and we subsequently work out the energy of the sparse vector, $\cb$, which is a compound random variable. By total expectation we have
\begin{eqnarray}
\mathbb{E}\left[\|\cb\|_{2}^{2}\right]=\mathbb{E}_{\,|\Ic_{c}|}\left[\mathbb{E}\left[\|\cb\|_{2}^{2}\big||\Ic_{\cb}|\right]\right]=\mathbb{E}_{\,|\Ic_{\cb}|}\left[|\Ic_{\cb}|\mathbb{E}\left[|c|^{2}\right]\right]=\mathbb{E}\left[|\Ic_{\cb}|\right]\mathbb{E}\left[|c|^{2}\right]\label{split}
\end{eqnarray}
where we have dropped the index in $\mathbb{E}\left[|c\,|^{2}\right]$ to denote an arbitrary nonzero
\emph{coefficient} of $\cb$. Using this result, we can see that
$\sigma_{\Cc}^{2}=\mathbb{E}[\|\cb\|_{2}^{2}]/N=\mathbb{E}[|c|^{2}]\mathbb{E}[|\Ic_{\cb}|]/N$
by Parseval's energy conservation law and an ergodicity assumption. Moreover,
\begin{eqnarray}\label{exp_c}
\mathbb{E}[|c|^{2}]=\mathbb{E}\left[|x|^{2}\big||x|>\gamma\right]-2\gamma\mathbb{E}\left[|x|\big||x|>\gamma\right]+\gamma^{2}
\end{eqnarray}
and so we derive the pdf of $|x|$ given $|x|>\gamma$ to be
\begin{eqnarray}\label{pdf_conditional} f\left(|x|\big||x|>\gamma\right)=\frac{f\left(|x|\right)\delta(|x|-\gamma)}{\bar{\mathbb{F}}^{2}_{|x|}(\gamma)}=\frac{|x|}{\sigma_{|x|}^{2}}e^{\frac{-|x|^{2}+\gamma^{2}}{2\sigma_{|x|}^{2}}}\delta(|x|-\gamma)
\end{eqnarray}
where $\bar{\mathbb{F}}_{|x|}(\gamma)=e^{\frac{-\gamma^{2}}{2\sigma_{|x|}^{2}}}$. Computing the terms in (\ref{exp_c}) using (\ref{pdf_conditional}) we get
\begin{eqnarray}
\mathbb{E}[|c|^{2}]=2\sigma_{|x|}^{2}-2\sqrt{\frac{\pi}{2}}\gamma\sigma_{|x|}e^{\frac{\gamma^{2}}{2\sigma_{|x|}^{2}}}\left(1-\textmd{erf}\left(\frac{\gamma}{\sqrt{2}\sigma_{|x|}}\right)\right)
\end{eqnarray}
Lastly, since $|\Ic_{\cb}|$ is a binomial, $\mathbb{E}[|\Ic_{\cb}|]=N\bar{\mathbb{F}}_{|x|}(\gamma)=N e^{\frac{-\gamma^{2}}{2\sigma_{|x|}^{2}}}$.
Substituting these last two expressions into (\ref{split}) gives $\sigma_{\Cc}^{2}$, which in turn produces (\ref{sigma_D}), and this parameter characterizes $f_{\Dc}$.

\section{Tailoring the Lambert $\mathcal{W}$-Function}\label{app2}
Given the equation $y=xe^{x}$, which for reasons stated below we call the \emph{canonical form}, it is desired to solve explicitly for $x$. Unfortunately, this could not be done using elementary operations. Instead, the solution can be expressed in terms of the Lambert $\Wc$-Function, where $x=\Wc(y)$ is the solution to the canonical form, and where we can thus equivalently write $y=\Wc(y)e^{\Wc(y)}$.

The function is generally multivalued. If we restrict its argument, $y$, to be real, then it produces two outputs for each point on the supporting interval $y\in[-e^{-1},0]$, which is our interval of interest. However, one of the two outputs of $\Wc(y)$ is $\geq -1$, and is referred to as the \emph{primary branch}, $\Wc_{0}(y)$, while the second is $<-1$, and is referred to as the secondary branch, $\Wc_{-1}(y)$. It will soon be clear that only the primary branch is needed (hence an injective mapping between $y$ and $x$ is retained). In any case, the function can be found iteratively by Newton's method; for instance, $x_{j+1}=x_{j}-\frac{x_j e^{x_{j}} - y}{e^{x_{j}}+x_j e^{x_{j}}}$. The problem at hand, as expressed in (\ref{concavity_condition}), is more complex than the canonical form. Nonetheless, it could be reduced to this form by a clever substitution \cite{Lambert}. First express (\ref{concavity_condition}) compactly as
\begin{eqnarray}
e^{a\tilde{r}+b}=c\tilde{r}+d,\label{abcd}
\end{eqnarray}
where $a=\sqrt{2}d_{\min}/\sigma_{D}^{2}$, $b=-d_{\min}^{2}/\sigma_{D}^{2}$, $c=\sqrt{2}d_{\min}/\sigma_{D}^{2}=a$, and $d=-1$. Letting $\rho=-a(r+\frac{d}{c})$ and substituting into (\ref{abcd}) gives $-\frac{a}{c}e^{-\frac{ad}{c}+b}=\rho e^{\rho}$. Comparing with the canonical form, the solution to the equation can be expressed as $\rho=\Wc_{0}(-\frac{a}{c}e^{-\frac{ad}{c}+b})$. Back-substitution to the four parameters in (\ref{abcd}) returns $\tilde{r}=-\frac{1}{a}\Wc_{0}\left(-\frac{a}{c}e^{-\frac{ad}{c}+b}\right)-\frac{d}{c}$. Returning the values of $a, b, c$ and $d$ into this equation gives the final expression in (\ref{Lambert_roots}).
\bibliographystyle{unsrt}
\bibliography{Reliability_SigPro_2014}

\end{document}